\documentclass[aps,prb,preprint,groupedaddress,amsmath,amssymb,showpacs]{revtex4}

\usepackage{epsfig}

\begin{document}

%\preprint{}


\title{
A global investigation of phase equilibria using the
Perturbed-Chain Statistical-Associating-Fluid-Theory (PC-SAFT)
approach}

\author{Leonid~Yelash$^a$, Marcus~M\"uller$^b$, Wolfgang~Paul$^a$, Kurt~Binder$^a$}

\affiliation{$^a$ Institute of Physics, WA 331, Johannes-Gutenberg University, D-55099 Mainz, Germany\\
$^b$ Department of Physics, University of Wisconsin, 1150 University Avenue, Madison, WI 53706}

\date{\today}

\begin{abstract}
\noindent The recently developed Perturbed-Chain Statistical
Associating Fluid Theory (PC-SAFT) is investigated for a wide
range of model parameters including the parameter $m$ representing
the chain length and the thermodynamic  temperature  $T$ and
pressure $p$. This approach is based upon the first-order
thermodynamic perturbation theory for chain molecules developed by
Wertheim [ M. S. Wertheim, J. Stat. Phys. {\bf 35}, 19--46 (1984);
{\em ibid.} {\bf 42}, 459--492 (1986)] and Chapman {\em et al.}
[G. Jackson, W. G. Chapman and K. E. Gubbins, Mol. Phys. {\bf 65},
1 (1988); W. G. Chapman, G. Jackson and K. E. Gubbins, {\em ibid.}
{\bf 65}, 1057 (1988)] and includes dispersion interactions via
the second-order perturbation theory of Barker and Henderson [J.
Chem. Phys. {\bf 47}, 4714 (1967)]. We systematically study a
hierarchy of models which are based on the PC-SAFT approach using
analytical model calculations and Monte Carlo simulations. For
one-component systems we find that the analytical model in
contrast to the simulation results exhibits two phase-separation
regions in addition to the common gas--liquid coexistence region:
One phase separation occurs at high density and low temperature.
The second demixing takes place at low density and high
temperature where usually the ideal gas phase is expected in the
phase diagram. These phenomena, which are referred to as
``liquid--liquid'' and ``gas--gas'' equilibria, give rise to
multiple critical points in one-component systems, as well as to
critical end points (CEP) and equilibria of three fluid phases,
which can usually be found in multicomponent mixtures only.
Furthermore, it is shown that the ``liquid--liquid'' demixing in
this  model is not a consequence of a ``softened'' repulsive
interaction as assumed in the theoretical derivation of the model.
Experimental data for the melt density of polybutadiene with
molecular mass $M_{\rm{w}}=45000$~g/mol are correlated here using
the PC-SAFT equation. It is shown that the discrepancies in
modeling  the polymer density at ambient temperature and high
pressure can be traced back to the ``liquid--liquid'' phase
separation predicted by the equation of state at low temperatures.
This investigation provides a basis for understanding possible
inaccuracies or even unexpected phase behavior which can occur in
engineering applications of the PC-SAFT model aiming at predicting
properties of macromolecular substances.
\end{abstract}




\maketitle


\section{Introduction}

The theoretical development of  analytical models to predict the
equation of state for polymer melts and solutions is a
longstanding problem of polymer science and very important for
engineering applications. As a consequence, much theoretical
effort has already been directed towards this problem. The
Flory--Huggins theory~\cite{flory}, for example, is based upon a
regular solution model which in a very simple manner considers
intermolecular interactions and chemical structure of  molecules.
This theory can  analytically describe generic, universal features
of polymer melts and solutions, e.g. the chain length dependence
of the critical properties and phase equilibria with an upper
critical solution temperature (UCST). Within this theory, however,
the third and all higher order virial coefficients exclusively
stem from the combinatorial entropy such that the monomer
interactions and liquid structure (packing) is only incompletely
described. Thus, the Flory-Huggins theory fails to describe, for
instance, a lower critical solution temperature (LCST) and does
not incorporate the compressibility (i.e., the pressure dependence
of the density) of mixtures. Another approach is based upon
equation-of-state description such as the Sanchez-Lacombe equation
\cite{sanchezlacombe,sanchezcho} which can parameterize the
compressibility of mixtures containing polymers.

In modern condensed matter science, however, one is dealing with
more complicated theories which are able to provide a very accurate
quantitative description of macromolecular systems.
Particular successful is an approach pioneered
by Wertheim~\cite{tpt1a,tpt1b,tpt1c,tpt1d,wh86,wh87} and
Chapman {\em et al.}~\cite{jchg88a,chjg88b}
based on liquid state theory.
In the classical form of perturbation theory of (simple) fluids,
one starts from the description of a
reference fluid with purely repulsive interactions for which the equation of
state and the pair correlation function, $g(r)$, are rather accurately
known.
Using the thermodynamic and structural properties of the repulsive reference
fluid as input, one treats the attractive interactions as a
perturbation \cite{zwanzig,bh1,bh2,wca}.
In contrast to this approach for simple fluids, the thermodynamic perturbation
theory for chain molecules (TPT1)
\cite{tpt1a,tpt1b,tpt1c,tpt1d,wh86,wh87,jchg88a,chjg88b}
employs the corresponding fluid of un-connected monomers
as starting point (a reference fluid),
and treats the chain connectivity as perturbation.

The first engineering application of this first order
thermodynamic perturbation theory has become well-known under the
name ``Statistical Associating Fluid Theory''
(SAFT)~\cite{chjgr89,chjgr90} and this name denotes any
implementation of the TPT1 method. This approach has been
extensively developed further \cite{muellergubbins01,economou02}
into several distinct variants. For example, using a square-well
potential of variable range (VR), the SAFT-VR method has been
proposed \cite{vrsaft1,vrsaft2} while the SAFT-LJ
\cite{kraska-gubbins1} and the soft-SAFT
\cite{softsaft1,softsaft2} methods use a Lennard-Jones fluid as
reference system. The TPT1-MSA method \cite{tptmsa2000,tptmsa2002}
also utilizes the Lennard-Jones fluid as a reference system which
can be analytically described within the mean spherical
approximation. Finally, the Perturbed-Chain SAFT (PC-SAFT) method
\cite{dissgross,pcsaft1,pcsaft2,pcsaft3} has been recently
proposed which is based upon a hard chain reference system
augmented by an attractive interaction derived from the Barker and
Henderson perturbation approach~\cite{zwanzig,bh1,bh2,wca}
extended to chain molecules with additional adjustment to
experimental data. This model has  attracted great attention due
to its industrial relevance as a modeling and predicting tool, for
example, for polar fluids (e.g. hydrofluoroethers) \cite{vijande},
polymer fractionation \cite{cheluget}, as well as a theoretically
sound equation of state for investigating barotropic phase
phenomena in binary fluid systems \cite{cisneros04} and predicting
the phase behavior in ternary and quaternary systems \cite{spuhl}.
Furthermore, a simplified version of the PC-SAFT model has been
proposed recently \cite{simplepcsaft} which is based on the
Carnahan-Starling equation for hard spheres \cite{cs}. Therefore,
it is not only of theoretical interest but also of practical
relevance to carry out a global investigation of the phase
behavior which can be predicted by the PC-SAFT approach, and to
test its accuracy by comparison with Monte Carlo simulations for
precisely the same model.

\section{Equation-of-State Model}

A detailed description and derivation of the original PC-SAFT
model can be found elsewhere \cite{dissgross,pcsaft1,pcsaft2}.
Here we briefly give an introduction into the PC-SAFT approach for
chain molecules.

In the perturbation approach \cite{zwanzig}, the Helmholtz free
energy of a system of molecules can be expressed as a sum of the
contributions from an unperturbed (reference) system where
particles only interact via repulsive forces and a perturbation
due to attractive interactions (dispersion forces):

\begin{equation}\label{eq1}
\frac {A} {Nk_{\rm{B}}T} =
\frac {A^{\rm{ref}}} {Nk_{\rm{B}}T} +
\frac {A^{\rm{disp}}} {Nk_{\rm{B}}T}
\end{equation}

\noindent
where ``ref'' and ``disp'' denote the contributions of
the unperturbed reference system
and the dispersion perturbation, respectively.

\subsection{Reference Hard-Chain Fluid}

In PC-SAFT, the reference, unperturbed system is the same as in SAFT:
It is a chain fluid composed of hard spheres. The Helmholtz free energy of the
hard-chain system
is given within the TPT1 formalism by:

\begin{equation}
\frac {A^{\rm{ref}}} {Nk_{\rm{B}}T} =
\frac {A^{\rm{id}}} {Nk_{\rm{B}}T} +
\bar{m} \frac {A^{\rm{hs}}} {Nk_{\rm{B}}T}+
\sum_{i} x_{i} (m_{i}-1)\frac {A^{\rm{chain}}_{i}} {Nk_{\rm{B}}T}
\end{equation}

\noindent
where $\bar{m}=\sum_{i} x_{i}m_{i}$ is an average number of segments
in a multi-component mixture of chain molecules; $m_{i}$ are the segment
numbers for the chains of component $i$.
For one-component chain fluids with
segment number $m$ this
expression simplifies to:

\begin{equation}\label{eqAref}
\frac {A^{\rm{ref}}} {Nk_{\rm{B}}T} =
\frac {A^{\rm{id}}} {Nk_{\rm{B}}T} +
m \frac {A^{\rm{hs}}} {Nk_{\rm{B}}T}+
(m-1) \frac {A^{\rm{chain}}} {Nk_{\rm{B}}T}
\end{equation}

\noindent Here $A^{\rm{id}}$ is the Helmholtz free energy of the
ideal gas system; $A^{\rm{hs}}$ is the residual hard-sphere
contribution due to the reduction of the system free volume by the
volume of monomers; $A^{\rm{chain}}$ is the contribution to the
Helmholtz free energy due to formation of bonds between the
monomers, which can be calculated from the contact value of the
pair correlation function of the reference hard sphere fluid. For
a multi-component mixtures one obtains:

\begin{equation}
\frac {A^{\rm{chain}}}{Nk_{\rm{B}}T}=-\ln g^{\rm{hs}}_{ij}
\end{equation}

\noindent The free energy, the compressibility factor, and the
contact pair correlation functions $g^{\rm{hs}}_{ij}$ for
components $i$ and $j$ of the hard-sphere mixture are modeled in
PC-SAFT using the BMCSL equation of state derived by Boubl\'ik
\cite{boublik70} and Mansoori et al. \cite{mansoori}:

\begin{equation}
\frac {A^{\rm{hs}}}{Nk_{\rm{B}}T}=\frac{1}{\zeta_0} \left[
\frac{3\zeta_1\zeta_2}{1-\zeta_3} +
\frac{\zeta_2^3}{\zeta_3(1-\zeta_3)^2}+
\left(\frac{\zeta_2^3}{\zeta_3^2}-\zeta_0\right)\ln (1-\zeta_3)
\right]
\end{equation}

\begin{equation}
Z^{\rm{hs}} =
\frac{1}{1-\zeta_3} +
\frac{3\zeta_1\zeta_2}{\zeta_0(1-\zeta_3)^2} +
\frac{3\zeta_2^3-\zeta_3\zeta_2^3}{\zeta_0(1-\zeta_3)^3}
\end{equation}

\begin{equation}\label{eqgij}
g_{ij}^{\rm{hs}} =
\frac{1}{1-\zeta_3} +
\left(\frac{d_i d_j}{d_i + d_j} \right) \frac{3\zeta_2}{(1-\zeta_3)^2} +
\left(\frac{d_i d_j}{d_i + d_j} \right)^2
\frac{2\zeta_2^2}{(1-\zeta_3)^3}
\end{equation}

\noindent
where

\begin{equation}\label{zeta}
\zeta_n = \frac{\pi}{6} \rho \sum_i x_i m_i d_i^n \;\;\;\; {\rm{for}} \;\; n=0,1,2,3
\end{equation}

\noindent
These equations have been derived in a similar way as the
Carnahan--Starling equation \cite{cs}, however starting
from the solutions of the Percus-Yevick integral equation
for mixtures of hard spheres. For a one-component system,
these equations simplify to the Carnahan-Starling hard-sphere equations:

\begin{equation}
\frac {A^{\rm{CS}}}{Nk_{\rm{B}}T}=
\frac{4\eta-3\eta^2}{(1-\eta)^2}
\end{equation}

\begin{equation}
Z^{\rm{CS}} =
\frac{1+\eta+\eta^2-\eta^3}{(1-\eta)^3}
\end{equation}

\begin{equation}
g_{\sigma}^{\rm{CS}} =
\frac{2-\eta}{2(1-\eta)^3}
\end{equation}

with the packing fraction

\begin{equation}\label{eqeta}
\eta = \frac{\pi}{6} \rho m d^3
\end{equation}

\subsection{Attractive Interactions}

The segments of the chain molecules can exert attractive forces
onto each other. This attraction acts between segments of the same
molecule as well as between different chains. Within the
perturbation approach, the Helmholtz free energy of a system with
attractive interactions can be expanded in a power series of the
inverse temperature, $\beta=1/(k_{\rm{B}}T)$, which is  the
high-temperature expansion. The exact Helmholtz free energy is
therefore described by the infinite sum of terms which represent
the contributions of different orders. In the infinite-temperature
limit, $\beta \to 0$, all the terms in this expansion vanish
except the zero-order term which corresponds to a reference fluid
without attractive interactions, i.e., an unperturbed system. If
this expansion is truncated after the term of order $\beta^2$, it
is called the second order perturbation theory, which for simple
fluids of spherical molecules with a hard-sphere reference fluid
has been derived by Barker and Henderson \cite{bh1,bh2}. In the
second-order approximation, the Helmholtz free energy of a
dispersion interaction is  given by:

\begin{equation}\label{eqAdisp}
\frac {A^{\rm{disp}}}{Nk_{\rm{B}}T}=
\frac {A_1}{Nk_{\rm{B}}T}+
\frac {A_2}{Nk_{\rm{B}}T}
\end{equation}

The first-order term, $A_1$,  in the perturbation expansion can be related to an
average of the number of segment pairs within
the range of the attractive interaction.
The higher-order terms are related to higher moments of the
distribution of this number.
Alder {\em et al.}~\cite{alder72} have investigated a fourth order
perturbation theory and shown that this quantity is nearly
Gaussian distributed.
If the distribution were  exactly normal,
the second order perturbation theory would be exact.
In the close-packing and the low-density limits,
the deviations from the normal distribution vanish.
Thus, one expects the second order perturbation theory
to be accurate at high density and low temperature,
provided that the above mentioned
distribution
is nearly normal and the perturbation terms well determined.

The perturbation contributions in PC-SAFT are expressed by
 equations which depend on the chain length parameter $m$
(differing from the perturbation theory of
spherical molecules):

\begin{equation}
\frac {A_1}{N k_{\rm{B}}T} =-2\pi\rho
\frac{\epsilon}{k_{\rm{B}}T} m^2\sigma^3 \; I_1(\rho,m)
\end{equation}

\begin{equation}\label{eqA2}
\frac {A_2}{N k_{\rm{B}}T} =-\pi\rho m
\left[ \frac{\partial }{\partial \rho} \rho (1+Z^{\rm{hc}})\right]^{-1}
\left(\frac{\epsilon}{N k_{\rm{B}}T}\right)^2 m^2\sigma^3 \; I_2(\rho,m)
\end{equation}

\noindent
where $Z^{\rm{hc}}$ is the residual compressibility factor of the reference
hard-chain fluid, which can  be calculated from
Eqs.~(\ref{eqAref})-(\ref{eqgij}).
The perturbation integrals,
$I_1(\rho,m)$ and $I_2(\rho,m)$, implicitly depend
on density and chain length parameter, $\rho$ and $m$.
Actually, these
integrals can be calculated from the pair correlation function,
which  in turn depends on
density and chain length:

\begin{equation}\label{I1x}
I_1(\rho,m) = \int_1^\infty \tilde{u}_1(x)\; g^{\rm{hc}}(x')\;x^2\; dx
\end{equation}

\begin{equation}\label{I2x}
I_2(\rho,m) = \int_1^\infty \tilde{u}_1(x)^2\; g^{\rm{hc}}(x')\;x^2\; dx
\end{equation}

\noindent
where $\tilde{u}_1=u_1/\epsilon$ is a normalized attraction potential;
$u_1$ is the attractive (perturbation) part of a potential;
$x=r/\sigma$ describes the distance between segments normalized by the
temperature-independent segment diameter $\sigma$.
The distance normalized by the temperature-dependent segment diameter is
denoted by
$x'=r/d(T)$.
$g^{\rm{hc}}(x)$ is the segment-segment pair correlation function of
hard-chain molecules.

The analytical evaluation of the integrals, $I_1(\rho,m)$ and
$I_2(\rho,m)$, which are required for the perturbation theory, is
difficult. Firstly, it requires analytical expressions for the
pair correlation functions of the hard-chain fluid that are
reliable over the entire range of density and chain length
parameter. These expressions can be obtained from
integral-equation theories for monomers and  short chains
\cite{chiew90, chiew91, tang-lu,  chang98, chang99}. However,
because of non-trivial correlations due to excluded volume
effects, such expressions become less accurate for long chains and
polymers, where for instance $m$ can be of order $10^3$ and
larger. Secondly, the integrals additionally  depend on the
interaction potential, $\tilde{u}_1$. For the  special case of
square-well interactions, which do not depend on the
intermolecular distance $r$ in the range of the attraction well
$\sigma<r<\lambda \sigma$, the integrals  simplify: They  depend
on the pair correlation function only and there is a simple
relation between the integrals:

\begin{equation}
I_2(\rho,m)= \frac{\partial}{\partial \rho} \left[\rho I_1(\rho,m)\right]
\end{equation}

\noindent For a general form of the intermolecular potential,
however, the perturbation integrals are non-trivial functions. A
numerical calculation of such integrals is often omitted in
engineering applications for practical reasons. In PC-SAFT one
follows an approach previously proposed for monomers
and dimers~\cite{gulati-hall, hino-prausnitz} and approximates
these integrals by a power series of sixth order in the packing
fraction:

\begin{equation}\label{eqI1}
I_1(\eta,m)= \sum_{i=0}^6 a_i(m) \eta^i
\end{equation}
\begin{equation}
I_2(\eta,m)= \sum_{i=0}^6 b_i(m) \eta^i
\end{equation}

\noindent where $\eta $ is the packing fraction, defined in
Eq.~(\ref{eqeta}). For chain molecules, the coefficients $a_i(m)$
and $b_i(m)$ depend on the number of segments $m$. This chain
length dependence is approximated by following expressions in
PC-SAFT:

\begin{equation}\label{eqai}
a_i(m)= a_{0i}+\frac{m-1}{m} a_{1i}+\frac{m-1}{m}\frac{m-2}{m} a_{2i}
\end{equation}
\begin{equation}\label{eqbi}
b_i(m)= b_{0i}+\frac{m-1}{m} b_{1i}+\frac{m-1}{m}\frac{m-2}{m} b_{2i}
\end{equation}

\noindent
Eqs.~(\ref{eq1}), (\ref{eqAref})--(\ref{eqA2}), and
(\ref{eqI1})--(\ref{eqbi})
define the PC-SAFT approach to chain molecules.
For equations Eqs.~(\ref{eqI1})--(\ref{eqbi}) one needs no less than
42 model constants, $a_{ij}$
and $b_{ij}$.
These are universal constants, in the sense that
these parameters
do not depend on a specific substance to be modeled, but
represent the properties of a certain class of
intermolecular potential function, e.g. the square-well potential, the
Lennard-Jones potential or real interactions.

Generally, Eqs.~(\ref{eqI1})--(\ref{eqbi}) represent a
two-dimensional frame with a set of constants, $a_{ij}$ and
$b_{ij}$, which can be fitted to any perturbation. In the course
of the derivation of the PC-SAFT approach, two sets of constants,
$a_{ij}$ and $b_{ij}$,  have been obtained by the fitting the
model to the square-well potential \cite{dissgross,swpcsaft}
(which is called SW-PC-SAFT here) or to real substances for
PC-SAFT \cite{dissgross,pcsaft1}. This is particularly useful
because on the one hand, substituting the parameter set, $a_{ij}$
and $b_{ij}$, one can use the PC-SAFT approach to describe the
behavior of a specific perturbation potential (e.g. the
square-well potential) and one can test the accuracy of this
perturbation approach by quantitative comparison with computer
simulations of exactly the same model. On the other hand, using
the parameter set, $a_{ij}$ and $b_{ij}$, obtained for real
substances, one can compare the model to Monte Carlo simulations
of a coarse-grained potential for chain molecules, e.g. the
``bead-spring'' model which has been successfully applied to model
real systems \cite{advances,peter04}.

\subsection{``Softened'' Repulsion}

The hard-sphere repulsion is a very simple approximation and it
only provides a
very crude description of repulsive interaction in real systems.
A more realistic
representation of real interactions is, for example,
the Lennard-Jones potential.
For modeling the Lennard-Jones repulsion,
Barker and Henderson~\cite{bh2} proposed an effective,
temperature-dependent diameter $d(T)$
which is based on the hard-sphere reference fluid diameter $\sigma$:

\begin{equation}\label{dt}
d(T) =
\int_{0}^{\sigma}\left[ 1-\exp \left( -\beta u(r) \right) \right] dr
\end{equation}

\noindent Using a pair potential $u(r)$ one can calculate from
Eq.~(\ref{dt}) the temperature-dependent diameter not only for a
truly soft potential, e.g. the Lennard-Jones potential, but also
for ``softened'' potentials like the Chen-Kreglewski potential
\cite{kreglewski}. The Chen-Kreglewski potential has an additional
repulsive step if compared to the square-well potential. It is
shown in Fig.~\ref{fig-ck}. Using the Barker-Henderson recipe for
the Chen-Kreglewski potential, one obtains a temperature-dependent
effective diameter \cite{kreglewski}:

\begin{equation}\label{dtck}
d(T) = \sigma \left[ 1-0.12 \exp\left( -3\epsilon\beta \right) \right]
\end{equation}

\noindent
At low temperature, the effective diameter approaches the hard-sphere
diameter $d(T\to 0) \to \sigma$.
In the high-temperature limit, the effective diameter is smaller than
the hard-sphere diameter $d(T\to \infty) \to 0.88\sigma$.
The temperature-dependence of the effective diameter for the
Chen-Kreglewski potential is shown in Fig.~\ref{fig-dt}.

In the PC-SAFT approach,
the hard-sphere repulsion is replaced by the ``softened''
repulsion of the Chen-Kreglewski potential,
which is modeled by substituting expression Eq.~(\ref{dtck}) for the
diameter $d$. This method has also been employed earlier for
the BACK \cite{kreglewski} equation and the SAFT \cite{huang-radosz}
equation.
In SW-PC-SAFT \cite{swpcsaft}, in contrast,
the repulsion is treated as for the hard-sphere fluid:
The segment diameter, $d=\sigma$, does not change with temperature.

\section{Different levels in the PC-SAFT approach}

It has been mentioned above that in the PC-SAFT approach to chain
molecules there are two approximations which describe the shape of
the interaction potential: i) The effective temperature-dependent
diameter of segments, $d(T)$, models the repulsive branch of the
potential curve; ii) The polynomial expansion  with a set of
universal constants, $a_{ij}$ and $b_{ij}$, approximates the
attractive part of the potential. These approximations introduce
into the model flexibility for adapting it to different
intermolecular potentials. Therefore, we can build here a
hierarchy of models which can be derived from the PC-SAFT
perturbation approach. The levels in this hierarchy correspond to
complexity of the interaction potential, if compared to the
reference-model potential. The level 0 represents the  reference
hard-chain fluid.

Level 1) \underline{SW-PC-SAFT}: The Square-Well Perturbed-Chain SAFT
is based on the square-well potential.
The universal constants, $a_{ij}$ and $b_{ij}$, for the perturbation
polynomials, $I_1(\rho,m)$ and $I_2(\rho,m)$, have
been obtained by fitting the model to the square-well potential
\cite{swpcsaft}.
The hard-sphere diameter, $d$, is constant in this model.

Level 2) \underline{CK-PC-SAFT}:
The Chen-Kreglewski Perturbed-Chain SAFT
is similar to  SW-PC-SAFT except that the Chen-Kreglewski repulsion is
modeled instead of the hard-sphere repulsion. As a consequence,
the effective,
temperature-dependent diameter given by Eq.~(\ref{dtck}) is substituted
for the diameter $d$.
In order to model the square-well attraction of the Chen-Kreglewski
potential, the same set of the universal constants, $a_{ij}$
and $b_{ij}$, is employed for the perturbation polynomials
as in SW-PC-SAFT.

Level 3) \underline{PC-SAFT}:
This model includes the effective
temperature-dependent diameter for the Chen-Kreglewski repulsion
(Eq.\ref{dtck}) and the universal constants, $a_{ij}$ and
$b_{ij}$,  obtained by fitting the model to real substances
\cite{pcsaft1}.

One particular advantage of this hierarchical approach is that not
only the final result -- the equation of state -- but also the
different approximations at various stages of the analytical
treatment can be evaluated. In the first two equation-of-state
models (levels 1 and 2), the interaction potential is well
defined: It is the square-well chain for the SW-PC-SAFT model and
the Chen-Kreglewski chain for the CK-PC-SAFT model. For these
models one can carry out Monte Carlo simulations of chain
molecules for exactly the same interaction potentials and thus
assess the quantitative accuracy of the perturbed-chain approach.
In the third model (PC-SAFT), only the ``softened'' repulsion is
well defined, whereas the dispersion interaction has resulted from
a fitting procedure to real substances. Thus, we additionally
perform Monte Carlo simulations of a  coarse-grained
``bead-spring'' model for chain molecules, which is able to
reproduce many properties of real substances
\cite{advances,peter04}. These simulation models and techniques
are described in the following part.

\subsection{Bead-Spring Model for Chain Molecules}

The bead-spring model for chain molecules is based upon the
shifted Lennard-Jones (LJ) and the Finitely Extensible Nonlinear
Elastic (FENE) potential \cite{kremergrest,advances,compphys2002}.
This model is schematically depicted in Fig.\ref{fig-ljfene}. The
Lennard-Jones potential describes the repulsion and dispersion
forces between the segments for both the intermolecular and
non-bonded intramolecular interactions:

\begin{equation}\label{eqlj}
U_{\rm{LJ}} (r) = 4\epsilon
\left[
\left(\frac{\sigma}{r}\right)^{12}-
\left(\frac{\sigma}{r}\right)^6
\right] + \frac{127}{4096} \epsilon
\end{equation}

\noindent

The Lennard-Jones potential has formally an infinite range. In
order to increases the efficiency of computer simulations, one can
reduce the interaction range of the potential. This does not
change the qualitative phase behavior and there are different ways
how to do that.\cite{SMIT} Here we use the cut-off distance at
$r_{\rm{cutoff}}=2
\sqrt[6]{2}\sigma$~~\cite{advances,compphys2002}, which is close
to the minimum of the pair correlation function beyond the second
correlation shell in the dense phase, thus minimizing the error of
cutoff. Furthermore, the potential is shifted in order to avoid
the energy discontinuity at the cut-off distance.

Additionally, the segments bonded along a chain molecule interact
via the FENE potential. In our coarse-grained bead-spring model,
the distance between the bonded segments can change to a certain
extent, e.g. upon variation of the pressure. There is an
equilibrium distribution of bond lengths for fixed thermodynamic
parameters. The stretching of the bonds is restricted by
introducing a bond-energy penalty, which diverges for the bond
length equal to $r=1.5\sigma$:

\begin{equation}\label{eqfene}
U_{\rm{FENE}} (r) = -33.75\epsilon \ln
\left[1- \left(\frac{r}{1.5\sigma}\right)^2\right]
\end{equation}

This choice of the constants in the FENE potential favors the
maxima of the distance distribution for the bonded and non-bonded
monomers at $r\approx 0.96\sigma$ and $r\approx 1.12\sigma$,
respectively. The repulsion between the bonded segments is modeled
by the Lennard-Jones potential. In combination, the Lennard-Jones
and the FENE potentials

\begin{equation}\label{eqljfene}
U_{\rm{LJ+FENE}} (r) = U_{\rm{LJ}} (r) + U_{\rm{FENE}} (r)
\end{equation}

\noindent exhibit a sharp minimum of the potential energy at an
equilibrium value of the bond length (see Fig.\ref{fig-ljfene}).

This bead-spring model is very well studied and has already been
applied to many different problems including, recently, the
coarse-grained modeling of polymer solutions~\cite{advances,
peter04}. It has been shown, for example, that this model predicts
phase behavior for mixtures of carbon dioxide/hexadecane which is
in good agreement with experimental data.

\subsection{Chen-Kreglewski Chain Molecules}

The ``softened'' repulsion of the Chen-Kreglewski potential is the
theoretical basis for the PC-SAFT equation. Therefore, it is
investigated here using computer simulations. The potential is
shown schematically in Fig.~\ref{fig-ck}. It is similar to the
square-well potential, however it has an additional repulsive step
(inverse well):

\begin{equation}\label{eqck}
U_{\rm{CK}} (r) = \left\{
\begin{array}{r@{\quad:\quad}l}
+\infty & r<\sigma-0.12\\
+3\epsilon & \sigma-0.12 < r < \sigma \\
-\epsilon & \sigma < r < \lambda \sigma \\
0 & \lambda \sigma <r
\end{array} \right.
\end{equation}

\noindent The height $+3\epsilon$ and the width $0.12\sigma$ of
the repulsion step were empirically determined in order to improve
the description of the equation of state for some small molecules,
e.g. noble gases and short alkanes \cite{kreglewski}. From a
comparison shown in Fig.~\ref{fig-ljck} one can see that the
Chen-Kreglewski potential is a rather crude approximation of
realistic interactions, e.g. the Lennard-Jones potential. The
interaction range of the Chen-Kreglewski potential is $1.5\sigma$,
which is by factor $\approx 1.5$ smaller than the cutoff distance
of the Lennard-Jones potential, if the range and energy
parameters, $\sigma$ and $\epsilon$, are taken the same in
$U_{\rm{LJ}}(r)$ and $U_{\rm{CK}}(r)$.

In order to perform the Monte Carlo simulations of the Chen-Kreglewski
potential and to compare directly with
the PC-SAFT approach, the computer code, which has been
developed in the Condensed
Matter Theory Group at the University of Mainz for the
Lennard-Jones+FENE model \cite{diss-virnau},
was generalized to different
interaction potentials.
For modeling chain molecules with the Chen-Kreglewski potential,
the monomers are connected by rigid bonds.
Monte Carlo simulations of precisely the same interaction potential
can therefore be carried out and quantitatively
compared to the analytical calculations with the CK-PC-SAFT model
in order to establish the accuracy of the perturbed-chain approach.
The CK-PC-SAFT equation represents level 2 in the
hierarchy of the PC-SAFT approach
described above.

\subsection{Square-Well Chain Molecules}

The square-well chain-molecule model
represents the level 1 in the hierarchy of the models derived
from the PC-SAFT
approach: It is a theoretical basis for
the SW-PC-SAFT equation \cite{swpcsaft}.
For a systematic investigation of this approach,
Monte Carlo simulations of
square-well chain molecules and monomers
are also included in our investigation.
In this model, the segments of chain molecules are connected by
the rigid bonds of length $\sigma$
and interact via a square-well potential of
the range $\lambda=1.5\sigma$.
Thus, the simulations are carried out for precisely the same
interaction potential as in the SW-PC-SAFT equation of state.

\section{Computational Details}

\subsection{Simulation Techniques}

Monte Carlo simulations are performed here with the $NPT$
simulation technique to obtain equation of state for chain
molecules and for monomers rather than to use the $NVT$ method or
to calculate phase coexistence from the Gibbs ensemble
simulations. The $NPT$ method can be used to investigate both the
equation of state as well as the first order transitions
\cite{frenkel-smit}, whereas in the $NVT$ simulations the system
can stay in a metastable state, e.g. at negative pressure.
However, $NPT$ simulations take longer than $NVT$ simulations due
to volume fluctuations.

The equation of state from the simulations can readily be compared
to the predictions of the PC-SAFT model. We did not systematically
study finite size effects in the simulations. Experience with
related models \cite{advances,peter04} has shown that finite size
effects are negligible except in the intermediate vicinity of
critical points. In the present paper, only data in the one-phase
region away from critical points are presented. Thus we do not
expect finite size effects to affect our conclusions.

The chain length $m=29$ beads per chain has been chosen because
such chains can be equilibrated in relatively short simulation
runs, and they are still long enough to represent the key features
of polymer molecules. It has also been chosen for a compatibility
reason to other extensive studies of a polybutadiene melt using
atomistic molecular dynamic simulations \cite{smith99,krushev}. In
the bead-spring model investigation, the simulation box contains
$160$ chain molecules. For the square-well and Chen-Kreglewski
models, 40 molecules per box are used. The simulations of monomers
($m=1$) are carried out using 1562 particles in the simulation
box. The initial configurations are generated using the
configurational bias Monte Carlo method \cite{frenkel-smit}. At
low and moderate density one can generate an initial configuration
by inserting the molecules into the simulation box until a desired
density is reached. At high density, the probability to
successfully insert a long chain molecule decreases rapidly.
However, one can start inserting the molecules into a larger box
and afterwards compress the box (e.g. in $NPT$ runs) in order to
increase the density. After generating an initial configuration,
it is equilibrated using periodic boundary conditions until the
system properties (e.g. density and energy) fluctuate around a
constant value. The equilibration time can vary largely depending
on the state parameters (temperature and density). At low density,
the equilibration time is usually very short and the system can be
brought into an equilibrium state after several MC steps. At high
density and low temperature, however, the relaxation processes
slow down rapidly and the equilibration time can be several
millions Monte Carlo steps. After reaching an equilibrium state,
$NPT$ simulations are performed in order to obtain good statistics
of measured quantities. We have also done a series of $NVT$
simulations for testing and compatibility reasons. However, the
pressure fluctuations in $NVT$ simulations can sometimes exceed
the value of the pressure, whilst the $NPT$ method provides high
accuracy of the measured density and there is no need for the
virial expression for the pressure.

In $NPT$ simulations, the volume of the simulation box is changed
by Monte Carlo moves at constant pressure and temperature.  In
addition to the volume change, the chain conformations are
modified by displacements which include local monomer moves and
reptation of the entire chain. In a local move, each monomer of a
molecule is tried once. In a reptation move, the monomer at one
molecule end, which is chosen randomly, is cut and attached to the
opposite end of the same molecule. This procedure is repeated for
all molecules in the simulation box. The number of moves in one
Monte Carlo step can be varied. The typical choice for the
simulations here is one volume change, one local move and 100
reptations in one Monte Carlo step. At high density ($\rho^{*}\ge
1$), the acceptance probability of the reptation moves can
decrease strongly; hence, the number of the reptation moves in one
Monte Carlo step is set to a lower value (10 or 1) in such cases.
For the same reason, the insertion of an entire chain molecule is
practically impossible at high densities.

\subsection{Calculating Phase Diagrams}

Using equation-of-state models, we calculate isotherms, spinodals,
phase equilibria and critical points. The pressure, $p$, (e.g. for
the isotherms) can be calculated in a straightforward way by
substituting the temperature, $T$, and the density, $\rho$, (or
the molar volume, $v$) into an equation of state. This calculation
method is free of errors (except numerical rounding errors). Phase
equilibria, spinodals and critical points are calculated by
iteratively searching for state points that fulfill the
corresponding thermodynamic conditions. After the convergence of
the iteration procedure, e.g. for the critical isotherms plotted
in the figures, we explicitly verified by substituting the results
in the equation of state or its derivatives that the solution
fulfills the required conditions (i.e., that the convergence error
is negligible).

The calculation of equilibrium of two phases at the same
temperature, $T$, and pressure, $p$, is equivalent to solving the
following set of two equations:

\begin{equation}\label{eq-equilibr}
\left .
\begin{array}{r@{ }l}
\mu(\rho {\rm '},T) - \mu(\rho {\rm ''},T) = 0 & \\
 & \\
p(\rho {\rm '},T) - p(\rho {\rm ''},T) =0 &
\end{array}
\right\}
\end{equation}

\noindent where $\rho {\rm '}$ and  $\rho {\rm ''}$ are the
densities of the coexisting phases being iterated to satisfy Eqs.
(\ref{eq-equilibr}). The pressure, $p$, is calculated from an
equation of state. The expression for the chemical potential,
$\mu$, can be derived from an equation of state using
thermodynamic relations. For a one-component system it is $\mu =
f+pv$, with $f=-\int p dv$ being the Helmholtz free energy.

Two-phase equilibria in one-component systems have one degree of
freedom ($d=1$) according to the Gibbs Phase Rule:

\begin{equation}\label{eq-gibbsrule}
d=c-ph+2
\end{equation}

\noindent where $c=1$ is the number of components and $ph=2$ the
number of phases at equilibrium. It follows that in Eqs.
(\ref{eq-equilibr}) one can freely vary only one of the variables:
$T$, $\rho {\rm '}$ and  $\rho {\rm ''}$. For example, one can
first fix the temperature, $T$, and obtain $\rho {\rm '}$ and
$\rho {\rm ''}$ via iteration of Eqs. (\ref{eq-equilibr}).

Spinodals are determined via the following thermodynamic condition:

\begin{equation}\label{eq-spin}
\left . \frac{\partial p(v,T)} {\partial v} \right | _T  = 0
\end{equation}

\noindent According to the Gibbs Phase Rule, spinodals have one
degree of freedom in one-component systems: one component ($c=1$)
and one phase ($ph=1$) from Eq. (\ref{eq-gibbsrule}) yield
$d=2$ which has to be reduced by one due to the spinodal condition
Eq. (\ref{eq-spin}), which is an additional condition reducing the
number of independent variables in Eq. (\ref{eq-gibbsrule}) from
two to one. Hence, one can first fix the temperature, $T$, and
iterate the density to satisfy the above condition. Alternatively,
one can iterate the temperature at constant density. Both methods
yield the values of $T$ and $\rho$, which can afterwards be
substituted into an equation of state to calculate the spinodal
pressure.

The calculation of critical points requires solving of two
equations (which can alternatively be also expressed in terms of
the density, $\rho$):

\begin{equation}\label{eq-crit}
\left . \frac{\partial p(v,T)} {\partial v} \right |_T  = \left . 
\frac{\partial^2
p(v,T)} {\partial v^2} \right |_T  =0
\end{equation}

\noindent From the Gibbs Phase Rule it follows that the critical
point is invariant in one component systems, i.e. it has zero
degree of freedom: $c=1$, $ph=1$ and two equations Eqs.
(\ref{eq-crit}) reducing the number of independent variables in
Eq. (\ref{eq-gibbsrule}) from two to zero yields $d=0$. Therefore,
both the temperature and the density must be iterated
simultaneously in Eqs.~(\ref{eq-crit}). The critical pressure is
calculated afterwards by substituting the results for $T$ and
$\rho$ into the equation of state.

For the spinodal and critical point conditions Eqs.
(\ref{eq-spin}) and (\ref{eq-crit}) one needs derivatives of the
pressure with respect to the volume. In our work here, we obtained
these derivatives analytically, which reduces the calculation
error strongly, if compared to a numerical evaluation of the
derivatives.

Eqs. (\ref{eq-equilibr}), (\ref{eq-spin}) and (\ref{eq-crit}) are
nonlinear equations, which preclude analytical evaluation of
solutions. Different iteration solvers can be used for such
problems, which however usually require initial (``guess'') values
of the variables. In the case of one-root problems, such initial
values can be taken far from an actual solution. However, if one
deals with a many-root problem, a good choice of initial variables
can be important. For example, if one would like to calculate the
gas--liquid and liquid--liquid equilibria  in binary mixtures or
the corresponding critical curves, one cannot use only one set of
initial parameters: These phenomena are usually separated from
each other in the phase diagram and, therefore, one needs
different initial parameters for solvers.

Similarly, one needs different initial parameters in order to
calculate all types of critical points, phase equilibria and
spinodals using the PC-SAFT model: gas--liquid-,
``liquid--liquid''- and ``gas--gas''-type of solutions exist
there. There are different ways how to obtain such initial values.
The simplest one is to scan, i.e. to calculate a series of
isotherms to find out where they change the type from super- to
subcritical and to use parameters in that vicinity for the
iterative solver to calculate the solution exactly. There is also
a more sophisticated method how to find out a new phenomenon in
the phase diagram. This method is based on the fact that often
there is a continuous transition between different critical/phase
phenomena, if one changes the molecular parameters, e.g. the chain
length parameter $m$ here or the energetic parameters
$\epsilon_{ij}$ for mixtures, rather than the thermodynamic-state
parameters. One can start, for example, from a known critical
state and trace that solution beyond a boundary state which is
usually a higher-order thermodynamic state such as a tricritical
point (TCP) in mixtures. At a boundary state, a phenomenon changes
its type, e.g. from a gas--liquid to a liquid--liquid type, and
one obtains parameters for the new state. The continuity concept
has been formulated for phase behavior in mixtures
\cite{schneider04} and frequently applied to global phase diagram
calculations \cite{yk-ber, yk-zeitschrift}.

Some misunderstanding can exist concerning the application of the
Gibbs Phase Rule to the multiple critical points and the
higher-order thermodynamic singularities, like double critical
point and critical end point, discussed for one-component systems
in this work. The Gibbs Phase Rule relating locally the
thermodynamic states to the degree of freedom gives no information
about the phase diagram topology globally. Since the critical
points located in different parts of the phase diagram are not in
equilibrium with each other, they comply with the Gibbs Phase Rule
independently. The higher-order thermodynamic states should have a
negative degree of freedom in experimental systems following the
Gibbs Phase Rule. For example, for a one-component critical end
point ($c=1$) with two phases in equilibrium ($ph=2$) and two
critical-point conditions Eqs. (\ref{eq-crit}) the Gibbs Phase
Rule Eqs. (\ref{eq-gibbsrule}) yields $d=-1$ for the degree of
freedom. However, such states are also thermodynamically
legitimate in theoretical investigations because an additional
variable, i.e. the segment parameter $m$, can be included into the
consideration increasing the number of independent variables in
Eqn.~(\ref{eq-gibbsrule}) from two to three. The critical end
point is invariant in this case, i.e. it can exist only for
isolated (discrete) values of the chain length parameter $m$ and
the thermodynamic variables $T$, $\rho$ and $p$. This discussion
is also true for a double critical point, for which the
thermodynamic relation is:

\begin{equation}\label{eq-dcp}
\left . \frac{\partial p(v,T)} {\partial v} \right  |_T  = 
\left . \frac{\partial^2 p(v,T)}{\partial v^2} \right |_T  = 
\left . \frac{\partial^3 p(v,T)} {\partial v^3} \right |_T =0
\end{equation}

\noindent The double critical point is a boundary state between
the metastable and unstable critical states, which looks like a
shoulder in a density-temperature diagram. One can compare a
double critical point and a critical end point in one-component
systems with a tricritical point in binary mixtures: It has the
degree of freedom $-1$ in binary mixtures. However, including
molecular parameters ($\epsilon_{ij}$) in addition to the
thermodynamic parameters in theoretical investigation makes the
degree of freedom non-negative for the tricritical point. This
method of the phase diagram investigation is called global phase
diagram method and used later on in this paper.

\section{Results}

\subsection{Phase Diagrams of Selected Systems}

In the following part, the investigation of the phase diagram is
reported for chain molecules with 29 beads per molecule and for
monomers using the different levels of the PC-SAFT approach
discussed above. The isotherms are compared to Monte Carlo
simulations of the corresponding molecular model. In fact, the
choice $m=29$ is convenient for Monte Carlo studies (considerably
longer chains would be difficult to equilibrate), and this choice
of $m$ is large enough, that properties for polymer melts are
expected.

In Fig.~\ref{fig-m29_pcsaft} the phase diagram is shown which has
been calculated using the PC-SAFT equation (level 3) for chain
molecules with $m=29$. The Monte Carlo simulations of the
bead-spring chain molecules with $m=29$ using both the $NPT$ and
$NVT$ ensembles are also included in the figure. The simulations
are performed for the same temperatures as the calculations with
PC-SAFT (except $T^*_{\rm c}=0.768$).

The equation of state predicts two coexistence regions: the
gas--liquid region at low density with reduced critical
temperature $T^*_{\rm{c}}=3.868$ and reduced critical pressure
$p^*_{\rm{c}}=0.00463$, and a ``liquid--liquid'' region at high
density with reduced critical temperature $T^*_{\rm{c}}=0.768$ and
reduced critical pressure $p^*_{\rm{c}}=5.66$. The reduced
parameters are given in Lennard-Jones units.

The gas--liquid critical temperature in the simulations is  much
lower than the equation-of-state critical temperature, which is
obvious from the shape of the isotherm at $T^*=3$: It is a
super-critical isotherm in the simulations, but it is a
liquid-type isotherm in the analytical model. From the shape of
the PC-SAFT isotherm at $T^*=1$ one can see that the influence of
the ``liquid--liquid'' demixing extends far into the region of the
homogeneous liquid: The super-critical isotherm flattens out. This
effect is very pronounced for $T^*=1$; therefore, it can be
expected for higher temperatures too.

In our Monte Carlo simulations of the coarse-grained bead-spring
model for chain molecules carried out down to temperature
$T^*=0.75$ we could not find any evidence of such
``liquid--liquid'' demixing. The pressure along the simulated
isotherms increases monotonously in marked contrast to the PC-SAFT
isotherms for $T^* < 1$.

At high pressure, the isotherms calculated with the analytical
model exhibit a crossing. The possibility that such thermodynamic
anomaly can occur in models employing a temperature-dependent
co-volume or volume translation method \cite{martin79} widely used
in the field of chemical engineering has previously been reported
in the literature~\cite{stb,pfohl,klgpd,aiche}. This region
however is not accessible to our Monte Carlo simulations.

Summarizing, the PC-SAFT equation exhibits unusual
``liquid--liquid'' demixing with a critical point and the crossing
of isotherms at high density. Recently, Magee and Wilding
\cite{magee-wilding} have reported a second critical point for the
Lennard-Jones-Devonshire cell model -- a basic cell model for the
Lennard-Jones fluid. Furthermore, White \cite{jwhite05} has
studied an analytical model for the potential with repulsive
shoulder and reported three critical points and closed--loop
liquid--liquid equilibria for one-component systems. Hence, it is
not for the first time that the analytical theories predict phase
demixing which does not exist in the underlying molecular model.
Our attempt to perform a quantitative comparison of the PC-SAFT
model and the Monte Carlo simulations of the coarse-grained
bead-spring model based on the Lennard-Jones+FENE potential does
not yield good agreement for the same choice of molecular
parameters. However, this disagreement might be due to inadequate
representation of the molecular interactions implicitly captured
by fitting the PC-SAFT equation to real substances. To investigate
the source of these discrepancies we now compare computer
simulations and corresponding analytical models.

In Fig.~\ref{fig-m29_ckpcsaft} the phase diagram for the
CK-PC-SAFT equation and Monte Carlo simulations of the
Chen-Kreglewski chain molecules are shown. This is precisely the
same potential as modeled by the CK-PC-SAFT equation. One can see
that the agreement between the equation of state and the
simulations is much better in this case (e.g. for the
high-temperature isotherms at $T^*=10$ and $T^*=4$ as well as for
the low-temperature isotherms $T^*=1$ and $T^*=0.75$. However, the
analytical calculations exhibit systematic deviations from the
simulation results for a liquid-like density: For temperatures
between $T^*=3$ and $T^*=1$, the density obtained from the
simulations is somewhat lower than that from the equation-of-state
calculations (i.e., for the same density the pressure from the
simulations is higher than from the equation of state). At high
($T^*\ge 4$) and low ($T^*\approx 1$) temperatures, however, these
deviations vanish. At low density and high temperature, the
situation is opposite: The pressure from the simulations is lower
than from the model calculations.

A remarkably good agreement is obtained for $T^*=3$ and $p^*=0.05$
in the vicinity of the gas--liquid critical point. The
equation-of-state critical point is at $T^*_{\rm{c}}=2.87$ and
$p^*_{\rm{c}}=0.03$. Typically, one would expect rather poor
agreement near a critical point, since it cannot be described by
mean field theories correctly. The critical fluctuations can be
accounted for by simulations, although they are restricted by the
size of the simulation box. However, they are completely ignored
in the equation-of-state models. A finite size scaling analysis
can be applied to computer simulations in order to determine the
critical point in the thermodynamic limit. Since we are interested
in gross feature of the equation-of-state model, which can also be
deduced from non-critical data, we set here aside the
investigation of the finite size analysis near the critical
points. The reason for this perfect agreement becomes clear if one
analyzes another systematic deviation between the simulations and
the equation-of-state calculations for isotherms $T^*=4$ and
$T^*=10$: The simulation results are above the calculated curves
at high pressure and are below these curves at low pressure.
Therefore, there is a pressure at which both methods yield similar
results. For $T^*=10$, this pressure is about $p^*\approx 1$, and
for $T^*=4$ it is $p^*\approx 0.2$ (see
Fig.~\ref{fig-m29_ckpcsaft}). For $T^*=3$, the same result is
obtained at $p^*=0.05$.

The ``liquid--liquid'' demixing found with PC-SAFT is also
predicted by CK-PC-SAFT, albeit at lower temperature: The
``liquid--liquid'' critical point is located at $T^*=0.2$. The
direct investigation of this region using Monte Carlo simulations
is not possible, since the chain molecules exhibit a glass
transition at high density ($\rho>1$), and the equilibration of
chains slows down very rapidly. Therefore, we base our
investigation on the further systematic analysis of the model
hierarchy. Moreover, the CK-PC-SAFT equation exhibits a crossing
of the isotherms at high pressure similar to the PC-SAFT equation.
A comparison of the simulation results to the analytical
calculations for $T^*=4$ and $T^*=10$ again shows a systematic
deviation as pressure increases, which is related to this crossing
of the isotherms.

Summarizing, the CK-PC-SAFT equation predicts the same topology of
the  phase diagram as the PC-SAFT equation with $m=29$: One finds
the crossing of the isotherms and the ``liquid--liquid'' phase
separation at high density but at much lower temperatures than in
PC-SAFT. The replacement of the PC-SAFT dispersion by the
square-well dispersion cannot correct these artifacts.

In the next step, we remove the approximation of the ``softened''
repulsion, modeled by the effective, temperature-dependent
diameter $d(T)$, which turns the Chen--Kreglewsi potential into
the square-well potential. In Fig.~\ref{fig-m29_swpcsaft} the
phase diagram for square-well chains with 29 beads/molecule
obtained using the SW-PC-SAFT equation and Monte Carlo simulations
of isotherms for precisely the same potential are shown. One can
observe that the removal of the effective-diameter approximation
avoids the crossing of the isotherms at high pressure and improves
the overall agreement between the equation-of-state calculations
and Monte Carlo simulations. Therefore, one can trace back the
crossing of the isotherms and the inaccuracy of the CK-PC-SAFT
equation at liquid density to the temperature-dependent diameter.

The critical temperature and the phase envelope of the
``liquid--liquid'' demixing are nearly identical for SW-PC-SAFT
and CK-PC-SAFT, because at low temperature the effective diameter,
$d(T)$,  approaches the hard-sphere diameter, $\sigma$ (see Fig.
\ref{fig-dt}). In fact, the value of $u(r)/k_{\rm{B}}T$ calculated
from the potential energy, $u(r)$, for the Chen-Kreglewski
``softened'' repulsion diverges at low temperature and in the
low-temperature limit it turns to be the same as for the
square-well potential (i.e. infinity).

Theoretical considerations suggest that liquid--liquid phase
coexistence can occur in systems with softened potentials
\cite{stell-hemmer} and this behavior was recently investigated in
computer simulations~\cite{jagla01, pellicane04}. However, the
``liquid--liquid'' demixing found with PC-SAFT and CK-PC-SAFT here
is also present in the square-well equation for chains
(SW-PC-SAFT) -- a model which does not include the ``softened''
repulsion,  and where the ``liquid--liquid'' demixing clearly is
not expected. Hence, one can conclude that the ``liquid--liquid''
demixing predicted by these equations of state is due to
inappropriate approximations in the dispersion term and does not
reflect the statistical mechanics of the model.

By setting $m=1$, one can reduce the PC-SAFT model of chain
molecules to monomers. In this case, the chain term in equation
Eq.(\ref{eqAref}) vanishes and the reference hard-chain fluid
turns into the hard-sphere fluid. The $m$ dependencies in the
dispersion term vanish too, and the set of universal constants
($a_{ij}$ and $b_{ij}$ with $i=0,1,2$ and $j=0..6$) for
Eqs.~(\ref{eqai}) and (\ref{eqbi}) reduces from 42 to 14 constants
($a_{0j}$ and $b_{0j}$). In Fig.~\ref{fig-zy} the compressibility
factor is shown as function of the packing fraction calculated
from the SW-PC-SAFT equation for monomers. At high temperatures,
the analytical results are in excellent agreement with the
simulations. The gas--liquid demixing occurs at low density and
high temperature. At low temperature and high density, the
SW-PC-SAFT equation for monomers predicts a ``liquid--liquid''
demixing with critical temperature $T^*_{\rm{c}}=0.296$. However,
this phase coexistence is not expected for the square-well
potential.

The PC-SAFT approach for a one-component system of monomers thus
exhibits unusual ``liquid--liquid'' phase separation phenomena,
which persists further for chain molecules. This demixing is not
related to the modeling of the ``softened'' repulsion, since this
repulsion is not present in the square-well potential, but rather
to the use of approximations for the dispersion term, e.g. a power
series with fitted coefficients. A related problem due to a power
series expansion of dispersion terms has already been reported in
the literature~\cite{koak}. In order to investigate this
problematic feature of the model systematically, we perform global
calculations of the phase behavior including the polymer-like long
chain molecules in our study.

\subsection{Global Phase Behavior}

For a systematic classification of the phase diagrams predicted
with the van der Waals equation of state for mixtures, van
Konynenburg and Scott \cite{scott70, scott80} have proposed a
method which is called global phase diagram method \cite{f-g,
g-f-d}. This method is based on  grouping the binary phase
diagrams according to a topological similarity of the phase
equilibrium surface in the space of the reduced molecular
(energetic and geometric) parameters. In binary mixtures, for
example, there are two sets of the molecular parameters for each
component, $(\epsilon_{11},\sigma_{11})$ and
$(\epsilon_{22},\sigma_{22})$, and one set of the
cross-interaction parameters, ($\epsilon_{12},\sigma_{12})$, which
are also called binary parameters. The different types of the
phase behavior are separated in the molecular parameter space by
the boundaries representing the higher order thermodynamic states
such as a tricritical point (TCP) and a double critical end point
(DCEP). The global phase diagram method has already been applied
to many equations of state, binary and ternary systems, polar and
polymeric substances
\cite{deiters-pegg,vanpelt91,kraska-deiters,deiters-bluma,kraska,
kolafa98,nezbeda99,yk-ber,p-bq,wang-wu-sadus,scott99,ykd}.

In a one-component chain fluid,
the effective chain interactions  of an entire
chain change for different chain lengths.
Thus, in contrast to simple molecules,
there is a variety of
one-component phase diagrams for different $m$ values and fixed
segment interaction parameters, $\epsilon$ and $\sigma$.
For real substances,
e.g., the homologous series of alkanes,
these phase diagrams are qualitatively similar; however,
they differ by the location in the thermodynamic parameter space:
The coexistence curve and critical point typically shift to lower
pressure and density and to higher temperature for long chains.
Within the PC-SAFT approach one can define a common set of
segment interaction
parameters, $\epsilon$ and $\sigma$, which is independent of
$m$, and study the topology of the phase diagrams
for different chain lengths.

In Fig.~\ref{fig-density1} the temperature-packing fraction phase
diagram of one-component chains calculated using the PC-SAFT
equation is shown. The chain length parameters included in these
calculations are $m=1$ and $m=100$. In addition to the physically
reasonable gas--liquid-type critical point, a
``liquid--liquid''-type demixing is found for all values of $m$ in
these calculations. As $m$ increases, the ``liquid--liquid''-type
phase separation shifts to lower packing fraction $\eta\approx
0.52$  and higher temperature $T^*\approx0.8$, a region of
temperature and density which is of experimental interest for
macromolecules. Furthermore, a ``gas--liquid--liquid''-type phase
coexistence can occur within the model. For monomers, this
three-phase equilibrium is found at low temperature and vanishing
pressure; for long chains the temperature increases to $T^*\approx
0.7$. The shape of the gas--liquid coexistence curve (and
spinodals) can hence be affected by these three-phase equilibria
due to the thermodynamic relations for the coexisting phases. For
example, the location of the phase in the middle (at $\eta\approx
0.5$) and the shape of the binodal are defined not only by the gas
phase at low density, but also by the second ``liquid'' phase at
high density.

A ``gas--gas''-type critical point can be found for PC-SAFT in the
low-density region of the phase diagram at packing fraction
$\eta<10^{-2}$ and chain length parameter $m \geq 70$. An
enlargement of this region is shown in Fig.~\ref{fig-density2} for
chain lengths $m=1,70,100,1000$. An additional critical point at
low density appears for chain length $m \approx 70$, indicated in
Fig.~\ref{fig-density2} by a shoulder on the left side of the
gas--liquid spinodal for $m=70$. For a larger $m$, this "gas--gas"
critical point touches the binodal, which yields first a critical
end point (CEP), and  then a "gas--gas--liquid" coexistence with
two stable critical points in the gas--liquid region of the phase
diagram. As an example, the case $m=100$ is shown in
Fig.~\ref{fig-density2}. For these and longer chains (e.g.,
$m=1000$ representing the polymer limit in the figure), this new
critical point affects the shape of spinodal and coexistence
curves. Thus, the PC-SAFT equation  erroneously describes the low
density behavior, where one would expect a nearly ideal-gas
behavior.

A global investigation of the pressure-temperature phase diagrams calculated
with PC-SAFT is shown in Fig.~\ref{fig-pt} and explained schematically
in Fig.~\ref{fig-tcp}.
For large $m$, the gas--liquid critical point can move to
negative pressure in PC-SAFT.
Usually, it is not possible that the gas phase coexists at negative
pressure. In contrast, liquids can exhibit a phase coexistence and
critical phenomena in a metastable state
at negative pressure \cite{imre02,imre04}.
The PC-SAFT model predicts a ``gas--gas''
demixing with a critical point for $m\ge 70$ as has been shown above.
For $m\approx 70$, a double critical point appears on the gaseous
spinodal,
which causes a cusp of the otherwise smooth spinodal curve
(Fig.~\ref{fig-pt} and inset 2 in Fig.~\ref{fig-tcp}).
This double critical point yields two new critical points:
one metastable ``gas--gas'' critical point and one
unstable critical point, which looks like a shoulder
on the spinodal curve in the
temperature-packing fraction diagram in Fig.~\ref{fig-density2}.

For $m$ slightly larger than $70$, the metastable ``gas--gas'' critical
point touches the binodal curve in a critical end point (CEP)
(inset 3 in Fig.~\ref{fig-tcp}).
For $m=100$, two stable critical points for the
``gas--gas''- and gas--liquid-type demixing
exist in the diagram as well as the ``gas--gas--liquid''-type
triple point with
three binodal curves in the vicinity (inset 4 in Fig.~\ref{fig-tcp}
and Fig.~\ref{fig-density2}).
Further increase of $m$ causes the gas--liquid critical point
to shift to lower pressure until it touches the binodal curve in
a second critical end point (CEP) (inset 5 in Fig.~\ref{fig-tcp}).
Beyond this CEP
the gas--liquid critical point becomes metastable and can shift to
negative pressure, e.g. for the case $m=1000$ shown in
Fig.~\ref{fig-pt} and  inset 6 in Fig.~\ref{fig-tcp}.
The new ``gas--gas'' critical point changes its type
to a gas--liquid critical point after passing the critical end
point and takes the place of the ``former'' gas--liquid critical point.
Such multiple critical phenomena and transformations are well known from
the investigations of the tricritical points in binary fluid
mixtures.
Here, however, they are found using the PC-SAFT
equation for one-component systems.

In contrast, the ``liquid--liquid'' demixing and the
``liquid--liquid'' critical point exist at physical pressure and
temperature for all values of the chain parameter $m$. The
``liquid--liquid'' critical pressure decreases slightly with
increasing chain length parameter $m$ and converges to a constant
value $p^*_{\rm{c}}\approx 5.194$ (in Lennard-Jones units) for
infinitely long polymers. The ``liquid--liquid'' critical
temperature converges to $T^*_{\rm{c}}\approx 0.796$ and the
coexistence region shifts to higher temperature.

\subsection{Application to Correlations of Polybutadiene Data}

In order to investigate the impact of the artificial phase
separations discussed above on applications of practical interest,
the PC-SAFT equation is applied to experimental data of
polybutadiene with molecular mass $M_{\rm{w}}=45000$~g/mol
provided by BASF Aktiengesellschaft. Experimental data for the
density of polybutadiene at various pressures and correlations
with the PC-SAFT model using two molecular parameter sets are
shown in Figure \ref{fig-pbdbasf}. The parameter set~1 with
$m=1800$ gives a very good representation of the liquid density at
high temperature. Some discrepancies can be found at low
temperature for high pressure isobars. The parameter set~2 with
$m=1098$ gives worse correlation of the experimental data. At high
pressure, these discrepancies are very large (note that the solid
curve denotes $p=200$~MPa and the dashed curve corresponds to
$p=100$~MPa in Fig.~\ref{fig-pbdbasf}). The calculated isobars
splay out at low temperature unlike the experimental isobars which
can be approximated with straight lines in this temperature region
fairly well. Furthermore, the  splay of the calculated isobars is
stronger at higher pressure (compare isobars  $200$~MPa and
$0.1~$MPa in Fig.~\ref{fig-pbdbasf}).

In Fig.~\ref{fig-pbd1912} the data are correlated by a new set of
parameters, which yields the best possible representation of the
experimental data. A comparison between the isobars for the lowest
and the highest pressures reveals that the density variation
decreases by about 30\% in that temperature range, i.e., the
temperature dependence of the isobaric density becomes weaker for
polybutadiene at high pressure. This can  be deduced from the
experimental data which approach each other at high pressure. For
the PC-SAFT model, however, the distance between the isotherms
increases at high density. This problem is caused by the unusual
``liquid--liquid'' phase separation region in PC-SAFT for the
parameters supposed to describe polybutadiene
shown in Fig.~\ref{fig-pbd1912}.

The flattening of the critical isotherm at the ``liquid--liquid''
critical point ($T_{\rm{c}}=214$~K, $p_{\rm{c}}=492$~MPa) leads to
deviations from the actual behavior already for isotherms at much
higher temperatures and much lower  pressures, since pressure
effects to this transition set in very gradually. Therefore, this
artifact of the PC-SAFT approach has a direct influence on the
correlation of the data for polybutadiene and excludes any valid
extrapolation of the polybutadiene behavior outside the
experimentally measured range.

\section{Conclusions}

The phase diagram of the Perturbed-Chain Statistical Associating
Fluid Theory (PC-SAFT) has been studied for a wide range of the
chain length parameter $m$ and thermodynamic parameters such as
temperature, pressure and density. For a systematic investigation,
a hierarchy of the models, which can be derived from the PC-SAFT
approach to chain molecules, has been analyzed and compared to
Monte Carlo simulations of the square-well potential, the
Chen-Kreglewski potential, and a coarse-grained bead-spring model
of chain molecules. The aim of this study was to test the ability
of such PC-SAFT based approaches to accurately predict equation of
state properties of polymers. Unfortunately, we have found that
this approach suffers from artificial ``gas--gas''
and ``liquid--liquid'' demixing phenomena.

The PC-SAFT equation predicts a ``gas--gas'' critical demixing at
low densities and a ``liquid--liquid'' critical demixing at high
densities in addition to the common gas--liquid demixing. These
critical phenomena and phase equilibria exist usually in mixtures,
however, within this theory they have been found for one-component
fluids. Based on a systematic analysis of a hierarchy of the
PC-SAFT approach it is shown that the ``liquid--liquid'' demixing
is not due to the ``softened'' repulsion, but rather caused by
approximations in the dispersion term of the equation of state. In
the following we summarize this rather unusual behavior.

The ``gas--gas'' critical point appears on the gaseous side of the
gas--liquid spinodal for the chain length parameter $m \approx 70$
and persists for larger $m$. Therefore, in the low-density part of
the phase diagram, one can find at high temperature two critical
points for the ``gas--gas'' and gas--liquid demixing, three-phase
equilibria (``gas--gas--liquid''), and two critical end points (a
gaseous critical phase in equilibrium with a liquid phase and a
gas phase in equilibrium with a gas--liquid-type critical phase).
For $m > 210$, the gas--liquid critical point shifts to negative
pressure and the ``gas--gas'' critical point becomes the
gas--liquid critical point. This transformation of the critical
point affects the shape of the coexistence curve.

The ``liquid--liquid'' phase separation has been found at high
density and low temperature where usually a solid-liquid phase
transition or a glass transition in polymers can be expected. At
high pressure, this ``liquid--liquid'' demixing exhibits a
critical point for all values of the chain length parameter $m$.
The location of the  ``liquid--liquid''-type critical point and
the  ``liquid--liquid'' immiscibility region in the phase diagram
depends on the chain length parameter. For long chains, the
``liquid--liquid'' demixing shifts to higher temperature and lower
pressure. In the limit of infinitely long chains, the parameters
of the ``liquid--liquid'' critical point approach finite values
for the reduced critical pressure $p^*_{\rm{c}}\approx 5.194$  and
the reduced critical temperature $T^*_{\rm{c}}\approx 0.796$ (in
Lennard-Jones units). These values correspond to about
$p_{\rm{c}}\approx 280$~MPa and $T_{\rm{c}}\approx 200$~K for
typical values of the Lennard-Jones parameters $\epsilon/k_{\rm
B}=250$~K and $\sigma=4$~\AA.

The thermodynamic state parameters that correspond to the
``gas--gas'' and ``liquid--liquid''  coexistence regions are quite
remote from those in typical experiments  with low and moderate
molecular weight polymers. Therefore at first sight one might
expect that in the region of experimental interest these
artificial phase separation phenomena, that do not correspond to
any physical behavior of the system, do not matter much.

However, for a polybutadiene melt, for example, it has been shown
here that the ability of the PC-SAFT equation to correlate the
density of macromolecular substances is affected by the region of
``liquid--liquid'' demixing. The immiscibility region in this case
is located at the packing fraction $\eta \ge 0.51$, whereas the
experimental polybutadiene data are in the range of the packing
fractions $0.44 \le \eta \le 0.49$. However, the
``liquid--liquid'' demixing extends its influence on the
thermodynamic properties of a homogeneous fluid far into the
experimentally relevant region of the state parameters. We recall
that mean field theory predicts a Curie--Weiss-type divergence of
the compressibility $\kappa \propto 1/(T-T_{\rm{c}})$ at the
critical point (with $T_{\rm{c}}$ being the critical temperature)
and, therefore, the compressibility is significantly enhanced over
a wide region in the temperature-density diagram in the vicinity
of the critical point.

Furthermore, the continuity concept formulated for phase behavior
of mixtures \cite{schneider04} if applied to one-component systems
means that a phase phenomenon existing in one region of the phase
diagram affects also  neighboring  regions. For example, an
isotherm  can change the slope because of a spinodal or a critical
point in its vicinity; a coexistence curve can exhibit a shoulder
due to a metastable critical point in the vicinity of a critical
end point (CEP). Furthermore, a wrong location of a spinodal can
predict the phase separation kinetics incorrectly. For caloric
properties and expansion coefficients one can expect large
deviations or even  divergence-like anomalies in the vicinity of
the unusual immiscibility regions since these quantities are
related to derivatives, e.g. the isobaric thermal expansion
$\alpha_P=V^{-1}(\partial V / \partial T)_P$ or the isothermal
compressibility $K_T=-V^{-1}(\partial V / \partial p)_T$.

The impact of the unusual phase behavior is also important when
the model is applied to high molecular weight polymers and
mixtures, e.g. polymer solutions or polymer blends. The mixing
rules  employed usually for modeling mixtures can cause the
critical points to shift into the experimentally relevant region
of thermodynamic parameters, especially for systems with large
differences between the critical parameters \cite{ilya02, ilya03}.
For polymers in such cases, one speaks about a hypothetical
critical point, which is related to the chain length parameter $m$
and the molecular parameters $\epsilon$ and $\sigma$. It is {\em a
priori} not clear whether this unusual phase separation moves
closer to the parameter region of physical interest, when a
solvent is added, or whether it moves further away. It seems
plausible that the answer to this question will not be universal,
but will rather depend on the nature of the solvent. These facts
have to be considered utilizing PC-SAFT models to predict
thermodynamic properties, e.g. extrapolations into regions which
cannot be fitted to the equation of state parameters due to the
lack of experimental data but which are of practical interest to
investigate.

\section{Acknowledgments}

This work was supported by  BASF Aktiengesellschaft (Ludwigshafen,
Germany). We benefited from stimulating discussions with H.~Weiss,
U.~Nieken,  W.~H\"ultenschmidt, Ch.~Ch.~Liew, A.~Moreira, O.~Evers
(BASF Aktiengesellschaft, Ludwigshafen). We are also grateful to
the BASF Aktiengesellschaft for supplying the thermodynamic data
on polybutadiene shown in Figs. \ref{fig-pbdbasf} and
\ref{fig-pbd1912} as well as the equation parameters and the
calculations shown in Fig. \ref{fig-pbdbasf}. We thank P.~Virnau
(MIT, Cambridge) for interesting discussions and for providing us
with the Monte Carlo LJ+FENE code as well as L. G.~MacDowell
(Madrid, Spain) for his fruitful collaboration. CPU time was
provided by the computing center of the Johannes-Gutenberg
University of Mainz.

\clearpage
\bibliography{literatur}
\bibliographystyle{apsrev}

\begin{widetext}

%Please compile a list of all figure captions on a separate page:
\clearpage
\begin{list}{}{\leftmargin 2cm \labelwidth 1.5cm \labelsep 0.5cm}

\item[\bf Fig. 1]
The interaction potential of Chen and Kreglewski \cite{kreglewski}.
The solid circles represent the segment diameter $\sigma$.
At that distance, the molecules repel each other with interaction
energy equal to $+3\epsilon$.
The interaction energy diverges at distance $r=0.88\sigma$ (dashed
circles).
For comparison, the interaction energy in the hard-sphere and
square-well models diverges at distance $r=\sigma$.
The areas between the solid circles and the dotted circles correspond
to the attraction well of the potential with energy depth $-\epsilon$.
The corresponding potential curve is shown on the right hand side.

\item[\bf Fig. 2]
Temperature-dependent diameter calculated by applying the Barker-Henderson
relation for the effective diameter to the Chen-Kreglewski potential.
The hard-sphere diameter $\sigma$ is recovered at low temperature.
In the high-temperature limit, the effective diameter is $d(T)=0.88\sigma$.
${T^*}=k_{\rm{B}}T / \epsilon$ is the reduced temperature.

\item[\bf Fig. 3]
Bead-spring model for chain molecules.
Circles represent the coarse-grained monomers.
Springs represent the interactions between the bonded monomers,
which interact via the Lennard-Jones+FENE potential.
Non-bonded monomers interact via the Lennard-Jones potential.
The corresponding potential curves are shown schematically
on the right hand side.

\item[\bf Fig. 4]
Approximation of the Lennard-Jones potential by the Chen-Kreglewski
potential with the same depth $\epsilon$ of the attraction well and
the segment diameter $\sigma$.

\item[\bf Fig. 5] Phase diagram and isotherms calculated for chain
molecules with $m=29$ beads per chain using the PC-SAFT equation
(curves). The equation-of-state model exhibits a gas--liquid
demixing at low density and a coexistence of two dense phases,
which is called ``liquid--liquid'' demixing here. The
gas--liquid-type reduced critical temperature and pressure are
$T^*_{\rm{c}}=3.868$ and $p^*_{\rm{c}}=0.00463$, respectively. The
``liquid--liquid''-type reduced critical temperature and pressure
are $T^*_{\rm{c}}=0.768$ and $p^*_{\rm{c}}=5.66$, respectively.
Bold curves are the coexistence regions; thin solid curves are the
calculated isotherms for different temperatures; dashed curves are
the critical isotherms. Monte Carlo simulations of the
Lennard-Jones+FENE chains with $m=29$ monomers per molecule
(symbols) are shown for the same temperatures as the isotherms
calculated with the equation of state (except
$T^{*}_{\rm{c}}=0.768$). Error bars show the standard deviation
for the density obtained in $NPT$ simulations and for the pressure
obtained in $NVT$ simulations. Dotted lines are guides to the eye
for the Monte Carlo isotherms. Reduced pressure, temperature, and
monomer density are given in the Lennard-Jones units: $p^* \equiv
p\sigma^3 / \epsilon$, $T^* \equiv k_B T  / \epsilon$, and $\rho^*
\equiv \rho\sigma^3$. The graph b) shows the vicinity of the
``liquid--liquid'' demixing predicted by the equation of state.
Note that the isotherms in the simulations (symbols) have a
different curvature as the subcritical ones in the analytical
model (curves).

\item[\bf Fig. 6] Phase diagram and isotherms calculated with the
CK-PC-SAFT equation (curves) and Monte Carlo simulations of
isotherms (symbols) for the Chen-Kreglewski chain molecules with
$m=29$ monomers per molecule. The temperatures for the
calculations and the simulations are
$T^*=10/4/3/2.87/2/1.6/1.4/1.2/1/0.75$. $T^*_{\rm{c}}=2.87$ is the
gas--liquid critical isotherm from CK-PC-SAFT. The
``liquid--liquid'' critical isotherm ($T^*_{\rm{c}}=0.2$) and one
sub-critical isotherm ($T^*=0.16$) calculated with the CK-PC-SAFT
equation are also shown. Dotted lines are guides to the eye for
the simulated isotherms. Reduced parameters are defined as in
Fig.~\ref{fig-m29_pcsaft}.

\item[\bf Fig. 7] Phase diagram and isotherms calculated with the
SW-PC-SAFT equation (curves) and Monte Carlo simulations of
isotherms (symbols) for the square-well chains with 29
beads/molecule. The calculated and simulated temperatures are
$T^*=4/3/2.5/2/1/0.75$. $T^*_{\rm{c}}=2.5$ is the
equation-of-state gas--liquid critical temperature. The
``liquid--liquid'' critical isotherm ($T^*_{\rm{c}}=0.2$) and one
sub-critical isotherm for the ``liquid--liquid'' region
($T^*=0.16$) calculated with the SW-PC-SAFT equation are also
shown. In SW-PC-SAFT, the segment diameter is
temperature-independent in contrast to the CK-PC-SAFT equation
shown in Fig.~\ref{fig-m29_ckpcsaft}. Dotted lines are guides to
the eye for the simulated isotherms. Reduced parameters are
defined as in Fig.~\ref{fig-m29_pcsaft}.

\item[\bf Fig. 8] Packing-fraction dependence of the
compressibility factor calculated with the SW-PC-SAFT equation for
monomers ($m=1$) for the reduced temperatures $T^*=3$, $2$, $1.5$,
$0.75$, and $0.296$. The isotherm $T^*_{\rm{c}}=0.296$ is the
critical isotherm for the ``liquid--liquid'' phase separation at
high density. The gas--liquid demixing exists at low packing
fraction and higher temperature. Symbols are Monte Carlo
simulations of the square-well monomers. Reduced temperature is
defined as in Fig.~\ref{fig-m29_pcsaft}.

\item[\bf Fig. 9] A temperature-packing fraction phase diagram of
one-component systems calculated using the PC-SAFT equation for
monomers ($m=1$) as well as for chain molecules with the chain
length parameter $m$=100. Solid curves denote binodals; dashed
curves correspond to spinodals. Dotted arrows show the
trajectories of the gas--liquid and ``liquid--liquid'' critical
points for different values of the chain length parameter $m$. The
reduced temperature is $T^*=k_BT / \epsilon$; the packing fraction
is $\eta=(\pi/6)\,\rho m d^3$.

\item[\bf Fig. 10] Enlarged low-density region of the phase
diagram in Fig.~\ref{fig-density1} showing "gas--gas" and
"gas--gas--liquid" equilibria. The values of $m$ are the numbers
of repeat units per chain molecule. Legend for curves as in
Fig.~\ref{fig-density1}.

\item[\bf Fig. 11] A one-component global phase diagram in
pressure-temperature coordinates calculated with the PC-SAFT
equation for different values of the chain length parameter $m$.
Dash-dotted curves are the gas--liquid and ``liquid--liquid''
binodals for the monomer; bold curve is the ``liquid--liquid''
binodal for a polymer; solid thin curves are spinodals. Open
circles are the gas--liquid and ``gas--gas'' critical points;
filled circles are ``liquid--liquid'' critical points of monomer
and polymer. Dotted curves show the trajectories of the
gas--liquid, ``gas--gas'' and ``liquid--liquid'' critical points
calculated for different values of the chain length parameter~$m$.

\item[\bf Fig. 12] Schematic explanation of the phase diagram
topology in the pressure--temperature plane calculated in
Fig.~\ref{fig-pt} using the PC-SAFT equation. The cases 1, 2, 4,
and 6 correspond to the calculations for the chain length
parameter $m=1$, $70$, $100$, and $1000$, respectively, shown in
Fig.~\ref{fig-pt}. The cases 3 and 5 correspond to $70<m<100$ and
$100<m<200$, respectively (not shown in Fig.~\ref{fig-pt}). For
$m\approx 210$, the gas--liquid critical point crosses the
boundary $p^*=0$ and moves to negative pressure. Legends: solid
curves are binodals (the ``liquid--liquid'' binodals shown in
insets 1 and 6 are left out in the other pictures for simplicity
of the presentation). Dashed curves are the gas--liquid and
``gas--gas'' spinodals. The ``liquid--liquid'' spinodals, which
enclose the ``liquid--liquid'' binodal, are omitted. Symbols
denote special thermodynamic states: critical points (CP) (open
circles); critical end points (CEP) (filled circles); double
critical point (DCP) (filled square); triple points (filled
triangles). This series of diagrams explains also how the
gas--liquid critical point calculated with PC-SAFT can move to
negative pressure for long chains (shown in inset~6).

\item[\bf Fig. 13]
Correlation of experimental data for polybutadiene
provided by BASF Aktiengesellschaft
using PC-SAFT.
Circles are polybutadiene data at pressures
0.1, 10, 50, 100, 150, 200~MPa
and temperatures
299.14,
317.34, 333.17, 349.08,
365.12, 381.15, 397.2, 413.32, 429.01, 445.27, 461.39~K.
Polybutadiene used in these measurements is characterized by
$M_{\rm{w}}=45000$~g/mol, $45$\% cis, $45$\% trans, $10$\% $1,2$ vinyl.
These data were correlated by two sets of molecular parameters.
Parameter set~1: $m/M_{\rm{w}}=0.04$, $m=1800$, $\sigma=3.467$~\AA,
$\epsilon/k_{\rm{B}} =280.3$~K (solid curves).
Parameter set~2: $m/M_{\rm{w}}=0.0244$, $m=1098$, $\sigma=4.065$~\AA,
$\epsilon/k_{\rm{B}}=257.3$~K (dashed curves). Only the isobars
$0.1$, $10$ and $100$~MPa
are shown for the parameter set~2.
These parameters are provided by BASF Aktiengesellschaft.
At low temperature, the calculated density diverges for both
sets of molecular parameters.
Triangles are the calculations by BASF Aktiengesellschaft for the
parameter set~1.

\item[\bf Fig. 14]
A comparison of experimental data (symbols) for polybutadiene melt
($M_{\rm{w}}=45000$~g/mol, BASF Aktiengesellschaft) and
calculations using the PC-SAFT equation of state (curves).
Polybutadiene data were measured at temperatures $T_{\rm{exp}}$:
299.14~K, 317.34~K, 333.17~K, 349.08~K,
365.12~K, 381.15~K, 397.2~K,
413.32~K, 429.01~K, 445.27~K, 461.39~K.
Calculations using PC-SAFT are shown for the experimental temperatures
$T_{\rm{exp}}$ as well as for
the critical ``liquid--liquid'' isotherm $T_{\rm{c}}=214$~K,
a super-critical isotherm at $T=240$~K and
a sub-critical isotherm at $T=200$~K.
PC-SAFT parameters are
$m/M_{\rm{w}}=0.04249$, $m=1912$, $\sigma=3.389$~\AA,
$\epsilon/k_{\rm{B}}=269.3$~K.
For the high-pressure data, a much too large
density is predicted (marked with a dashed circle) as a result of
the vicinity of the spurious ``liquid--liquid''
phase separation predicted by the equation of state.
This phase separation
(although located at high density and low temperature) affects
the correlation and prediction of the experimental
data due to
increasing intervals between the isotherms (shown with arrows).
The packing fraction values (numbers in brackets) are shown for
the $x$-axis
in addition to the density values.
These numbers are approximate
since the exact values of the packing fraction and the density are
related by the diameter, $d(T)$, which is temperature dependent.

\end{list}

\clearpage

\begin{figure}[ht]
\begin{center}
\includegraphics[height=1.7in, width=3.3in, angle=-0]
{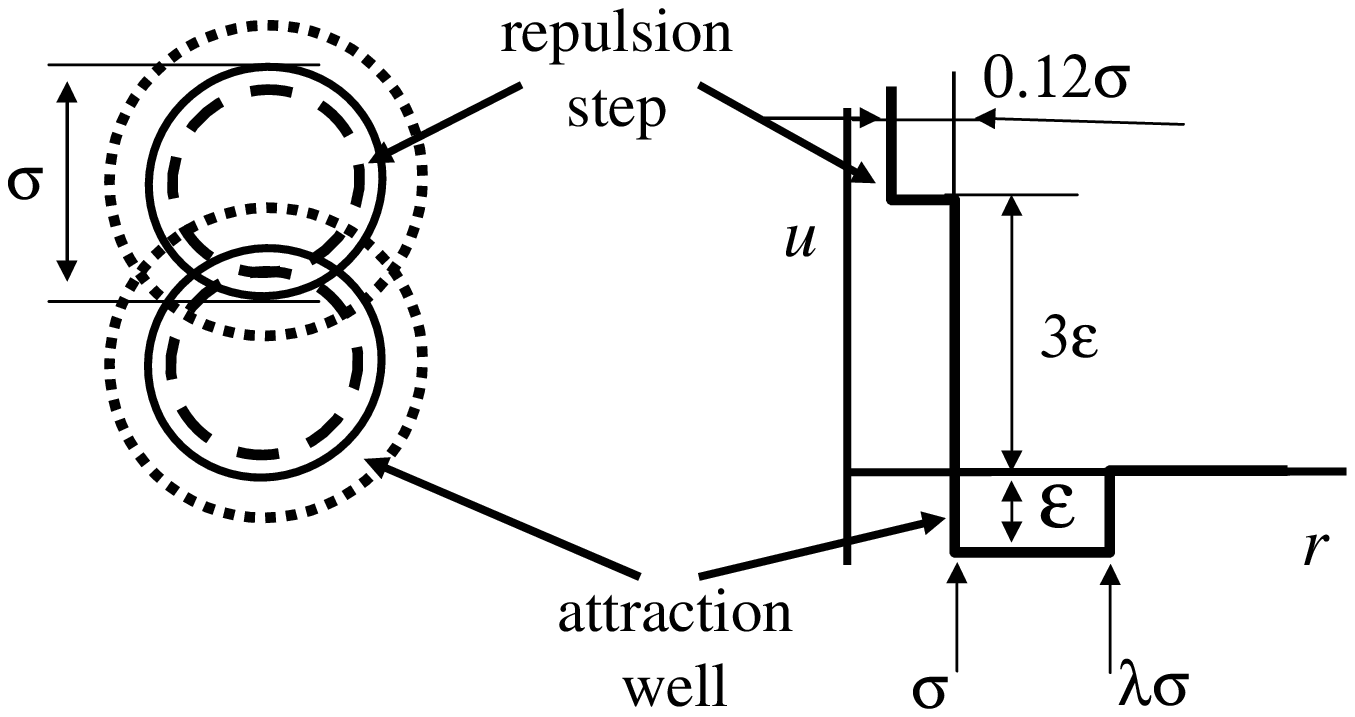}
\caption{} \label{fig-ck}
\end{center}
\end{figure}

\clearpage

\begin{figure}[ht]
\begin{center}
\includegraphics[height=2.3in, width=3.3in, angle=-0]
{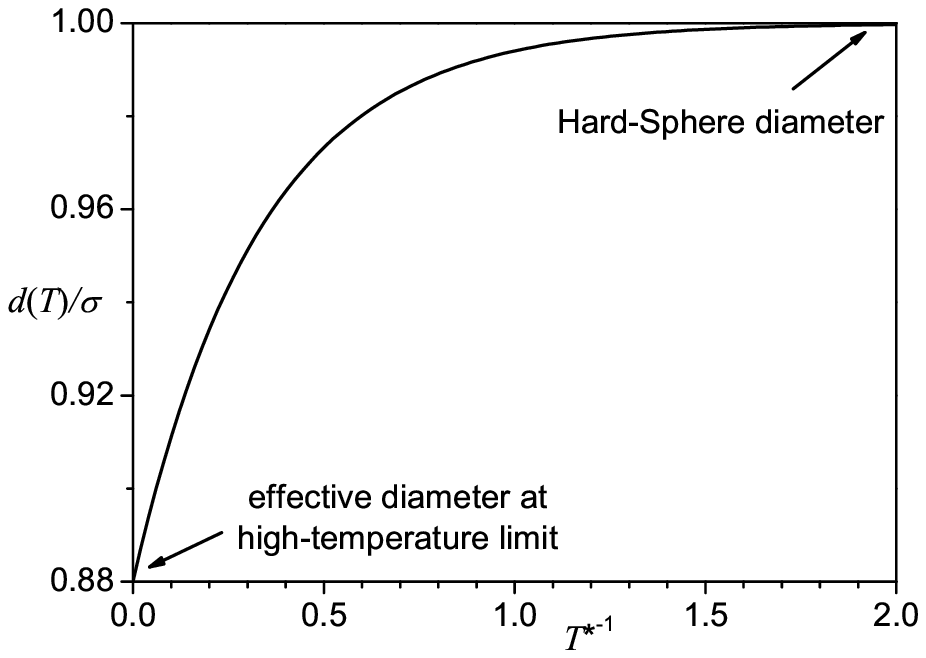}
\caption{} \label{fig-dt}
\end{center}
\end{figure}

\clearpage

\begin{figure}[ht]
\begin{center}
\includegraphics[height=1.5in, width=3.3in, angle=-0]
{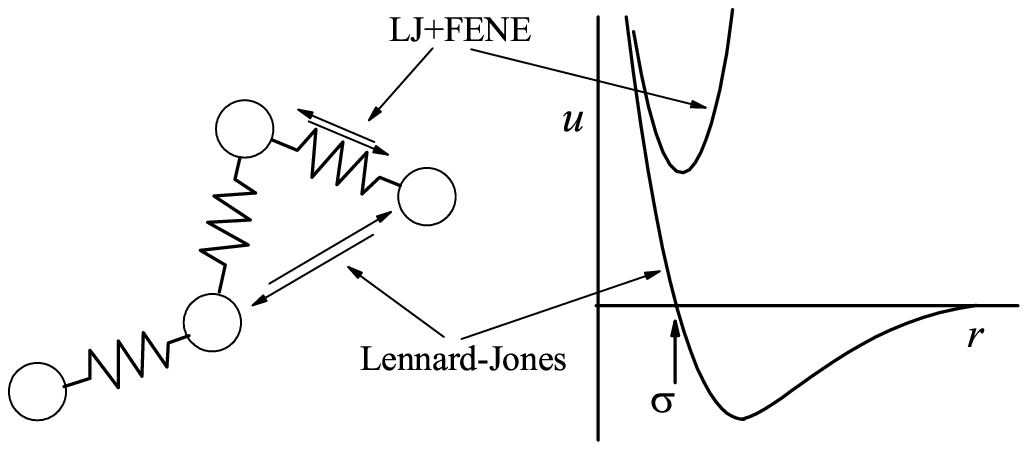}
\caption{} \label{fig-ljfene}
\end{center}
\end{figure}

\clearpage

\begin{figure}[ht]
\begin{center}
\includegraphics[height=2.3in, width=3.3in, angle=-0]
{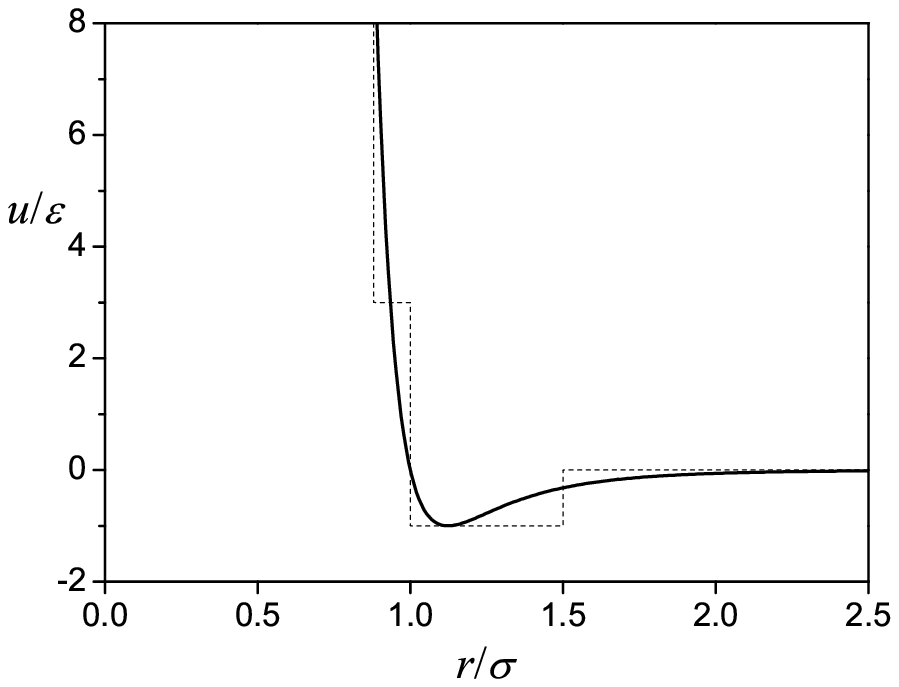}
\caption{} \label{fig-ljck}
\end{center}
\end{figure}

\clearpage

\begin{figure}[ht]
\begin{center}
\includegraphics[height=2.5in, width=3.3in, angle=-0]
{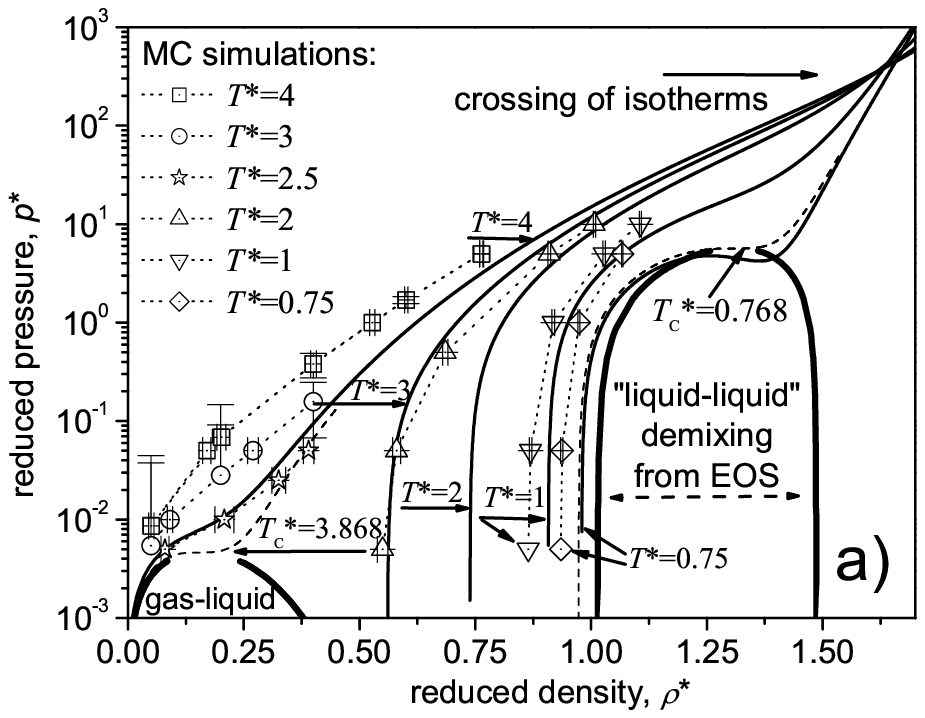}
\includegraphics[height=2.5in, width=3.3in, angle=-0]
{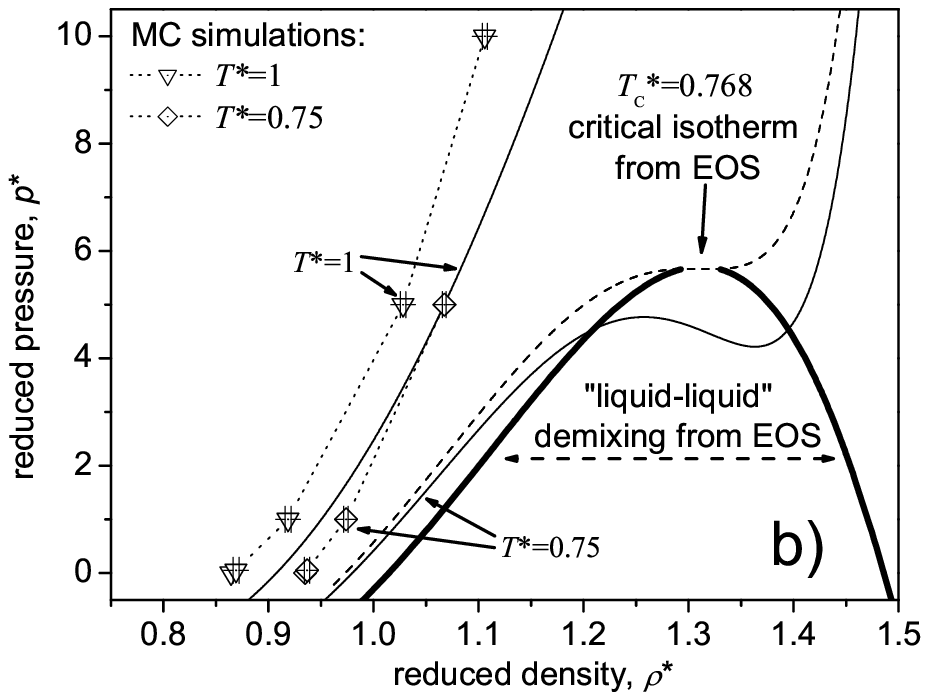}
\caption{} \label{fig-m29_pcsaft}
\end{center}
\end{figure}

\clearpage

\begin{figure}[ht]
\begin{center}
\includegraphics[height=2.5in, width=3.3in, angle=-0]
{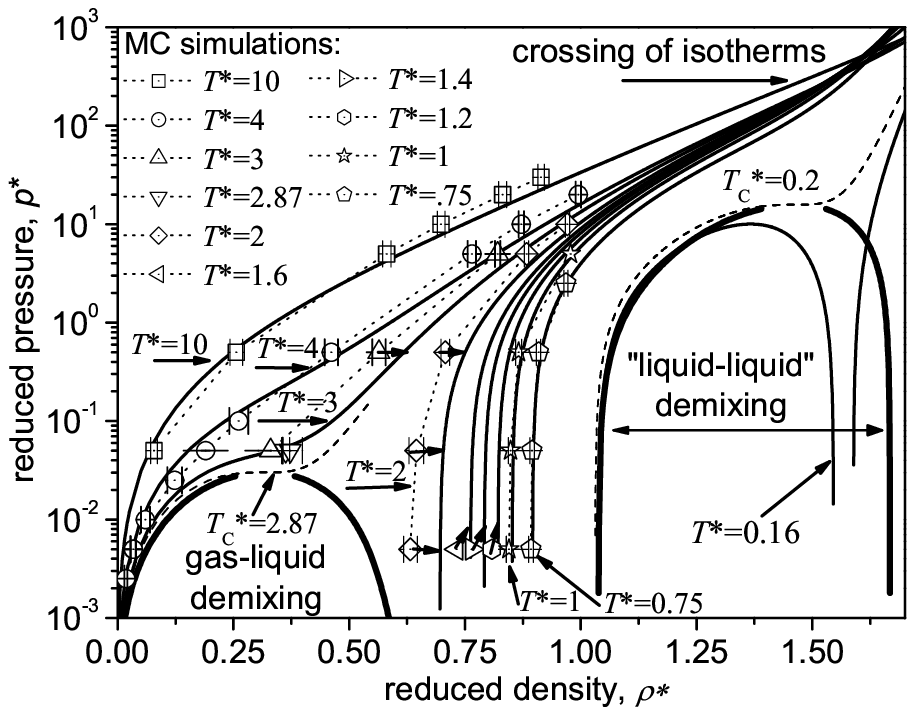}
\caption{}\label{fig-m29_ckpcsaft}
\end{center}
\end{figure}

\clearpage

\begin{figure}[ht]
\begin{center}
\includegraphics[height=2.5in, width=3.3in, angle=-0]
{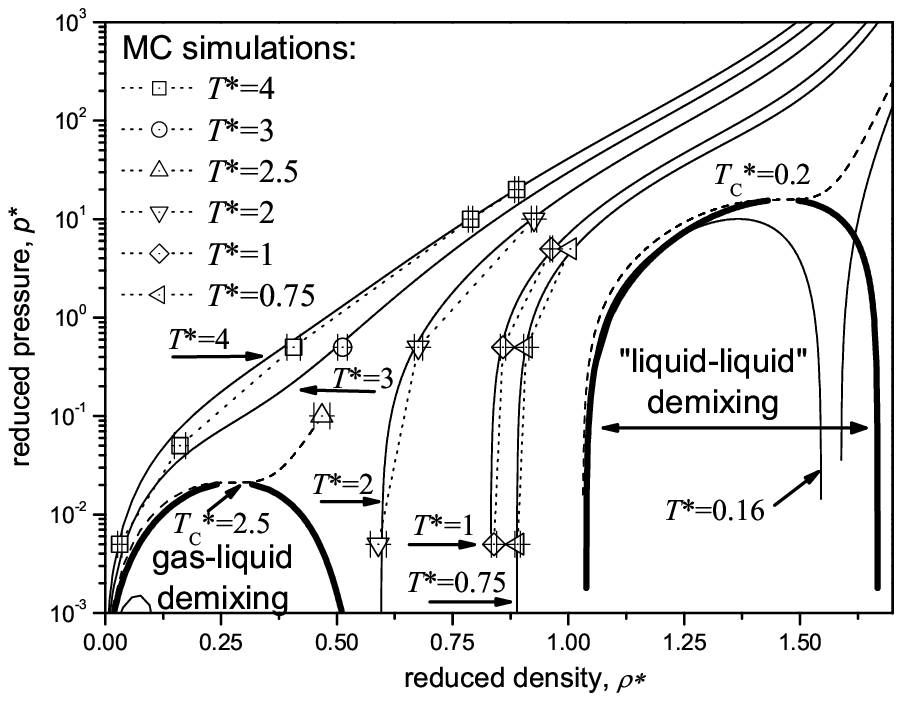}
\caption{}\label{fig-m29_swpcsaft}
\end{center}
\end{figure}

\clearpage

\begin{figure}[ht]
\begin{center}
\includegraphics[height=2.5in, width=3.3in, angle=-0]
{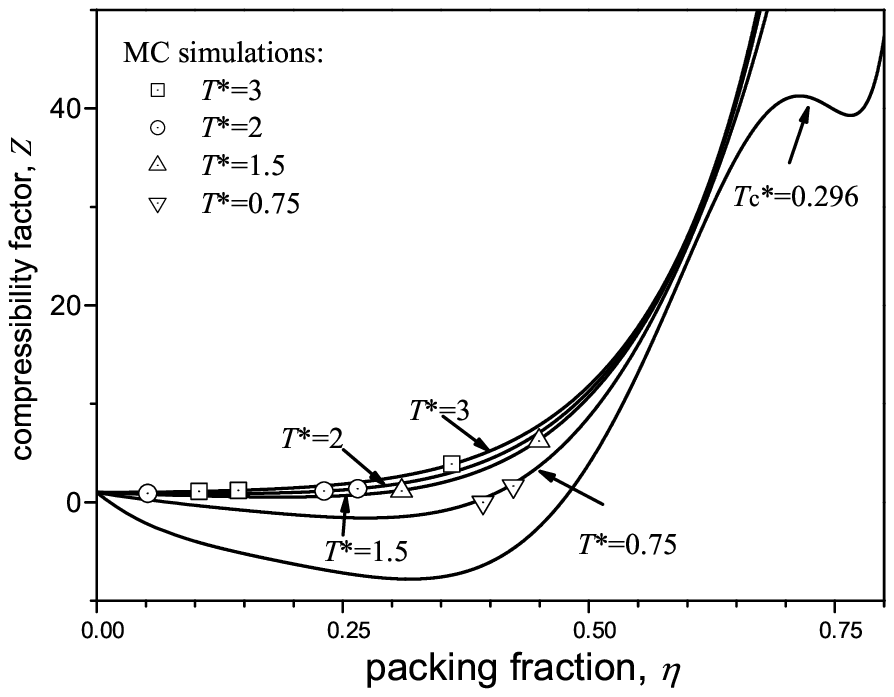}
\caption{}\label{fig-zy}
\end{center}
\end{figure}

\clearpage

\begin{figure}[ht]
\begin{center}
\includegraphics[height=2.5in, width=3.3in, angle=-0]
{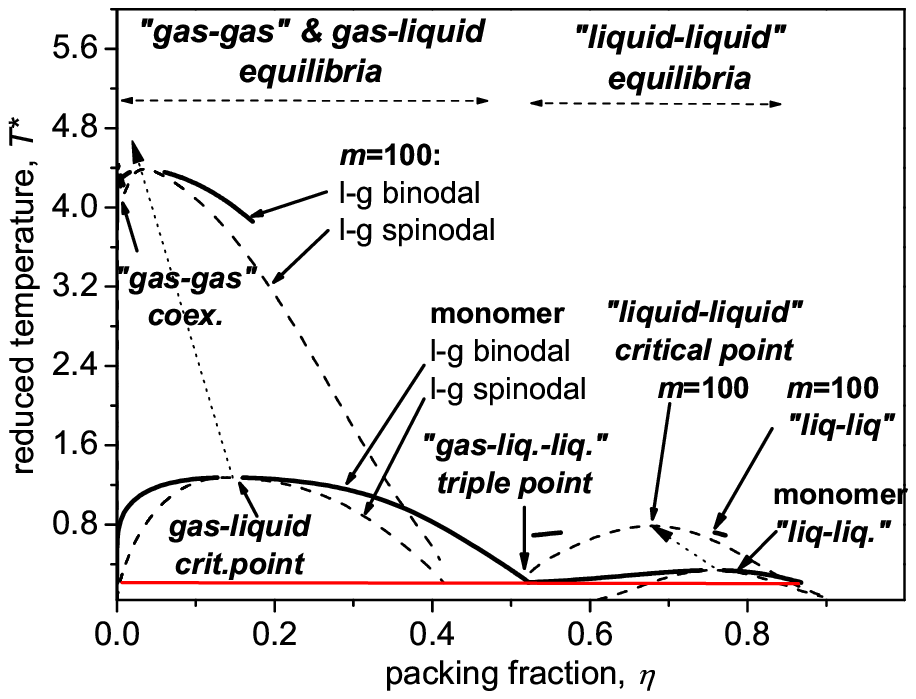}
\caption{} \label{fig-density1}
\end{center}
\end{figure}

\clearpage

\begin{figure}[ht]
\begin{center}
\includegraphics[height=2.5in, width=3.3in, angle=-0]
{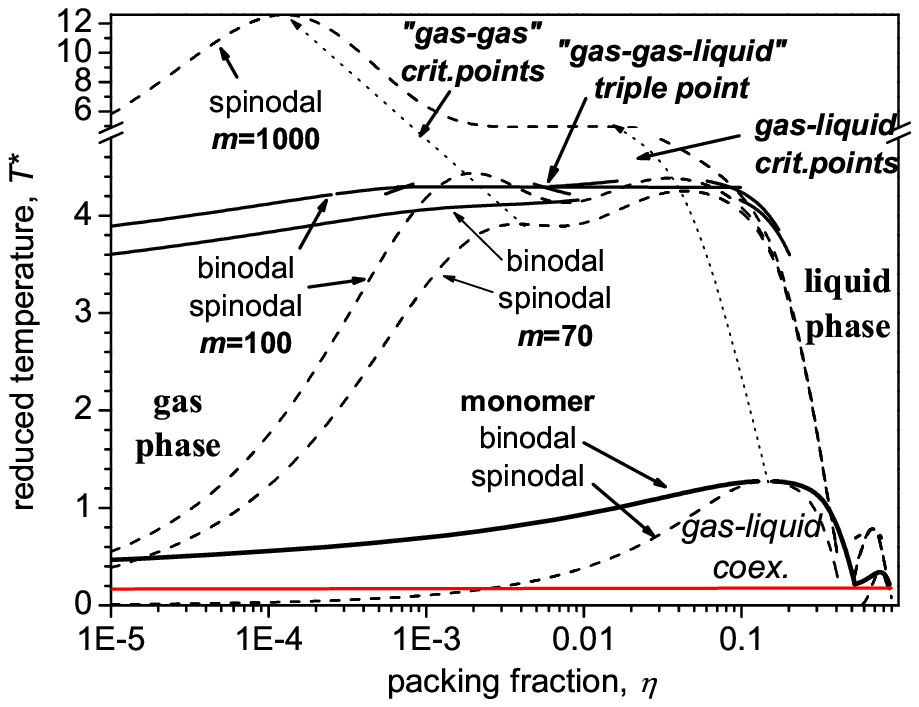}
\caption{} \label{fig-density2}
\end{center}
\end{figure}

\clearpage

\begin{figure}[ht]
\begin{center}
\includegraphics[height=2.5in, width=3.3in, angle=-0]
{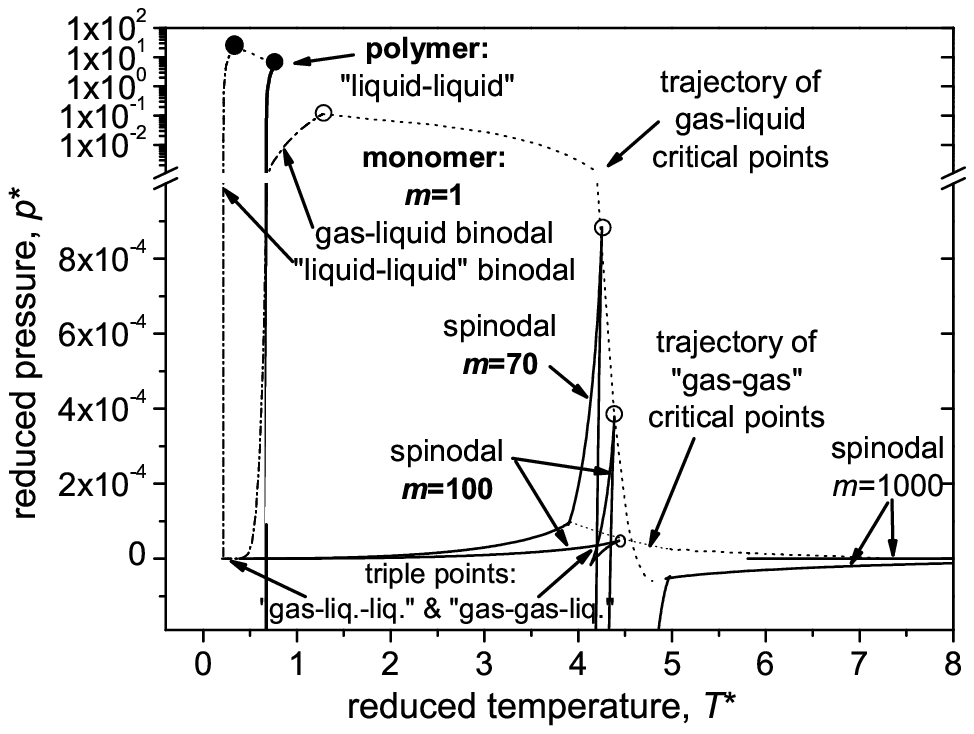}
\caption{} \label{fig-pt}
\end{center}
\end{figure}

\clearpage

\begin{figure}[ht]
\begin{center}
\includegraphics[height=2.5in, width=3.3in, angle=-0]
{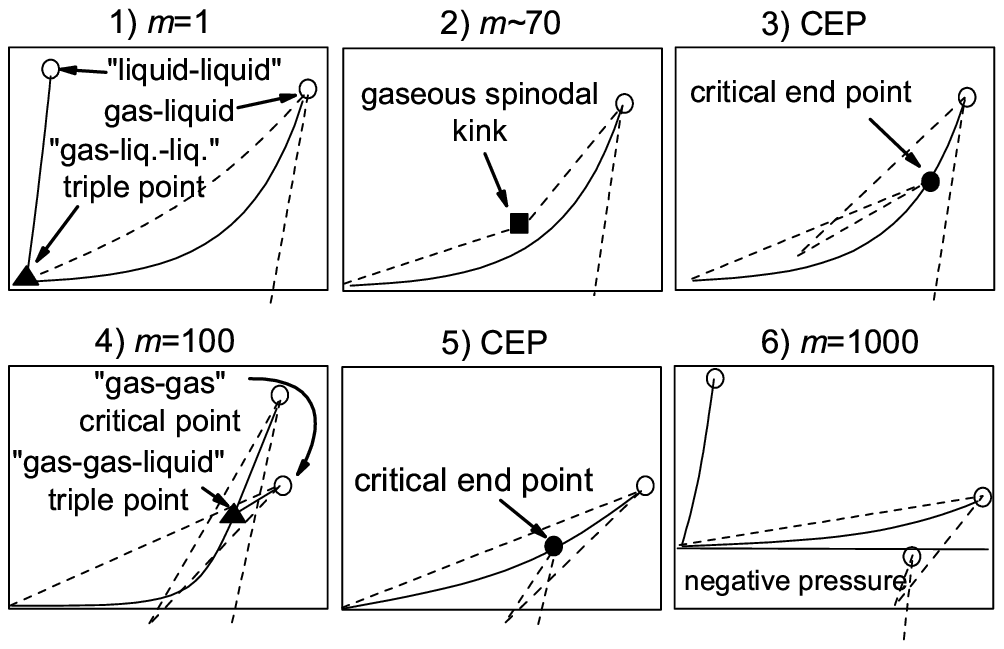}
\caption{} \label{fig-tcp}
\end{center}
\end{figure}

\clearpage

\begin{figure}[ht]
\begin{center}
\includegraphics[height=2.5in, width=3.3in, angle=-0]
{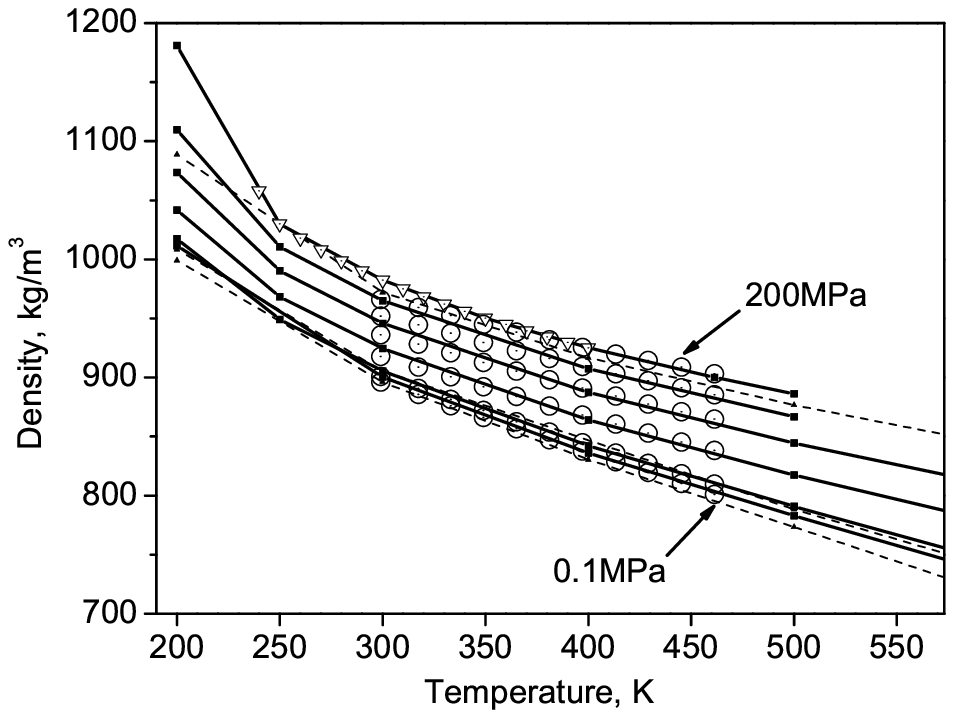}
\caption{} \label{fig-pbdbasf}
\end{center}
\end{figure}

\clearpage

\begin{figure}[ht]
\begin{center}
\includegraphics[height=2.5in, width=3.3in, angle=-0]
{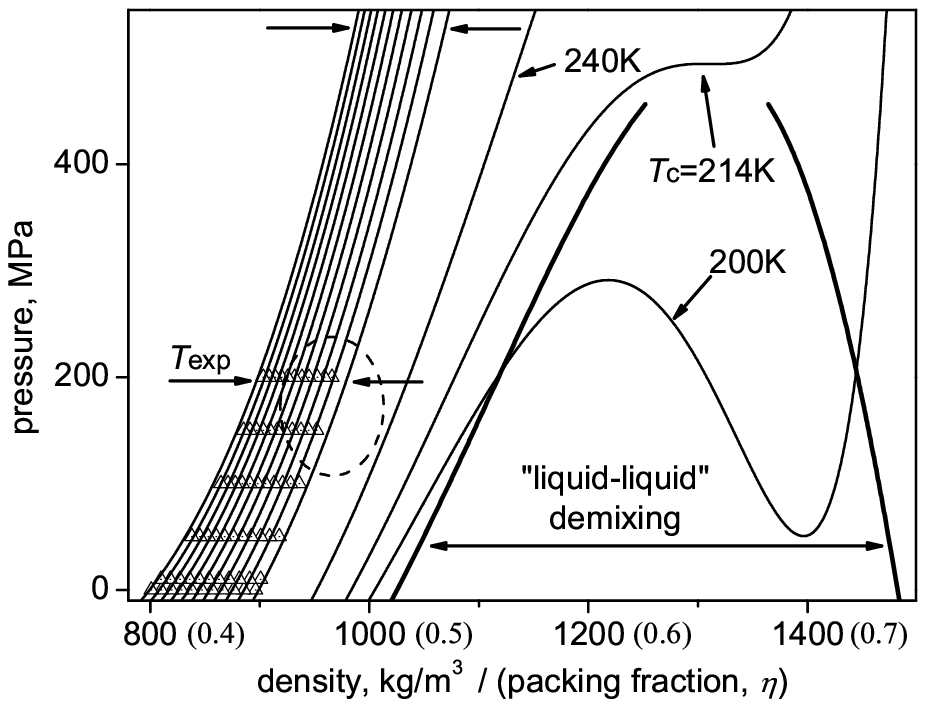}
\caption{} \label{fig-pbd1912}
\end{center}
\end{figure}

\end{widetext}
\end{document}